\documentclass[11pt,a4paper]{article}
\usepackage{jheppub,amsmath,amssymb,slashed,url,bm,textgreek,upgreek}
\usepackage{jheppub}  
\usepackage{tikz,lipsum,lmodern}
\usepackage[most]{tcolorbox}
\usepackage{amssymb} 
\usepackage{amsmath}
\usepackage{mathtools}
\usepackage{amsfonts}    
\usepackage{dsfont}
\usepackage{pdfpages}
\usepackage{verbatim}
\usepackage{tensor}
\usepackage{mathrsfs}
\usepackage{textgreek} 
\usepackage[mathscr]{euscript}
\usepackage[normalem]{ulem}
\usepackage{tikz}
\usepackage{makecell}
\usepackage{caption}
\usepackage{subcaption}
\usetikzlibrary{3d, arrows.meta, decorations.pathreplacing, decorations.markings,calc,shapes.misc,decorations.pathmorphing,patterns.meta, math}

\newcommand{\beq}{\begin{equation}}
\newcommand{\eeq}{\end{equation}}
 
\newcommand{\bea}{\begin{eqnarray}}
\newcommand{\ea}{\end{eqnarray}}
\newcommand{\barr}{\begin{array}}
\newcommand{\earr}{\end{array}}

\def\ie{\begin{equation}\begin{aligned}}
\def\fe{\end{aligned}\end{equation}}

\def\d{{\rm d}}
\def\i{{\rm i}}

\newcommand\PSU{\text{PSU}}
\newcommand\U{\text{U}}

\newcommand\SL{\text{SL}}
\newcommand\SU{\text{SU}}
\newcommand\SO{\text{SO}}

\newcommand{\cN}{{\cal N}}

\title{Spectrum of BPS black holes in $AdS_3 \times S^3 \times S^3 \times S^1$
\vspace{-0.5cm}
}

 \author{Matthew Heydeman${}^1$, Xiaoyi Shi${}^2$, Gustavo J. Turiaci${}^2$}
 \affiliation{${}^1$ Department of Physics \& The Black Hole Initiative, Harvard University, Cambridge, MA, USA}
 \affiliation{${}^2$ Physics Department, University of Washington, Seattle, WA, USA}\emailAdd{mheydeman@fas.harvard.edu, xiaoys5@uw.edu, turiaci@uw.edu}
\abstract{We uncover novel features in the spectrum of BPS and near-BPS states in asymptotically $AdS_3 \times S^3 \times S^3 \times S^1$ spacetimes. This follows from a careful analysis of semiclassical and quantum black holes in this theory, which have peculiarities due to the nonlinear large $\mathcal{N}=4$ superconformal symmetry. Notably, we find that the $S^3 \times S^3$ angular momentum spectrum of BPS states in the Ramond sector exhibits discrete jumps as a function of the ratio between the radii of the two three-spheres. This phenomenon is a quantum gravity effect for which no microscopic derivation is currently known. In addition, we construct a family of non-extremal supersymmetric black holes that contribute to a supersymmetric index yet possess a temperature-dependent free energy. Analogous results apply to six-dimensional black holes with $AdS_2 \times S^2 \times S^2$ near-horizon geometries constructed in M-theory compactifications.  }
\begin{document}\maketitle

\thispagestyle{empty} 

\setcounter{page}{1}

\section{Introduction} 
Understanding the full microscopic structure and dynamics of quantum black holes remains a central challenge in theoretical physics. A fruitful strategy has been to analyze supersymmetric (BPS) black holes in supergravity or string theory, where certain protected quantities may be computed in a weakly coupled or holographic description. Typically, the more supersymmetries, the more exact information can be extracted. In~\cite{Heydeman:2025vcc}, we utilized the Schwarzian effective theory (describing the near extremal perturbations of the black hole~\cite{Nayak:2018qej,Moitra:2018jqs,Castro:2018ffi,Sachdev:2019bjn,Ghosh:2019rcj,Iliesiu:2020qvm,Heydeman:2020hhw,Boruch:2022tno,Iliesiu:2022onk,Choi:2023syx,Rakic:2023vhv,Kapec:2023ruw,Kolanowski:2024zrq,Heydeman:2024ezi,Heydeman:2024fgk,Maulik:2024dwq,Maulik:2025hax,Brown:2024ajk,Lin:2025wof}) to classify supersymmetric black holes. We showed that black holes which preserve more than four supercharges generically lead to pathologies in the path integral.

Much of the logic for why the Schwarzian theory gives useful results for higher-dimensional BPS black holes is explained in \cite{Heydeman:2020hhw}. The main example considered there are black holes preserving 4 supercharges arising in flat 4d space, or in $AdS_3$ $\times$ $S^3$ $\times$ $M$ solutions of Type IIB supergravity, with $M = T^4$ or $K3$. This is dual to the D1-D5 system with ``small'' $\mathcal{N}=(4,4)$ superconformal symmetry~\cite{Maldacena:1997re,Strominger:1996sh,Horowitz:1996ay,Maldacena:1998bw}. 
 While the background is  10-dimensional string theory, many essential IR features are well captured by effective $AdS_3$ supergravity. For instance, black holes in this setting~\cite{Banados:1992wn,Banados:1992gq,Coussaert:1993jp} offer valuable insight and capture the exponential growth of the CFT density of states at high energies. If we instead take the near-extremal (low temperature) limit, one can still use modular invariance to derive the spectrum of near-extremal black holes, reproducing the predictions from the Schwarzian theory \cite{Ghosh:2019rcj}. Applying this logic~\cite{Heydeman:2020hhw} in the AdS$_3$ $\times$ $S^3$ $\times$ $M$ case leads to predictions for the BPS and near-BPS black hole spectrum from the $\mathcal{N}=4$ Schwarzian  based on the supergroup $\PSU(1,1|2)$. Many of these results have also been recently understood in \cite{Ferko:2024uxi} from the worldsheet point of view. 

An intriguing extension of this case arises in $AdS_3 \times S_+^3 \times S_-^3 \times S^1$ backgrounds of Type IIB string theory~\cite{Cowdall:1998bu,Boonstra:1998yu,Gauntlett:1998kc,Elitzur:1998mm,deBoer:1999gea}. The $\pm$ labels are introduced simply to distinguish the two inequivalent 3-spheres.  This background is constructed out of $Q_1$ D1-branes intersecting with $Q_5^+$ and $Q_5^-$ D5-branes and fluxes in type IIB string theory, and it has ``large'' $\mathcal{N}=(4,4)$ superconformal symmetry (denoted $A_\gamma$)  whose global supergroup is D$(2,1|\upalpha)$. Here, the supercharges are charged under both $\SU(2)_\pm$ factors of the $SO(4)$ R-symmetry, contrasting with the single $\SU(2)$ in small $\mathcal{N}=4$. There are two $\SU(2)_\pm$ current algebras with levels $k_\pm = Q_1 Q_5^\pm$, and the parameter $\upalpha = Q_5^-/Q_5^+ = k_-/k_+$ controls the ratio of radii of the $S_\pm^3$. A Schwarzian theory based on this supergroup was studied in \cite{Heydeman:2025vcc} as a particularly interesting case, and a central goal of the present work is to use these results to make detailed predictions for the spectrum of black holes in $AdS_3 \times S_+^3 \times S_-^3 \times S^1$\footnote{While our work was in preparation, we became aware of \cite{MR_WOP} which considers this background, including the supersymmetric black holes, from the worldsheet point of view. Their work is complementary to ours, and we achieve consistent results.}.

One reason for studying the black hole spectrum in this context is that surprisingly little is known about the microscopic theory. Despite the apparent similarity to the D1-D5 example, efforts to identify the ``D1-D5$^+$-D5$^-$ CFT'' have encountered challenges~\cite{Gukov:2004ym,Gukov:2004fh,Tong:2014yna,Gaberdiel:2013vva,Borsato:2015mma}, due in part to the nonlinear BPS bounds in $A_\gamma$, where the conformal weight depends in a complicated way on the $\SU(2)_+ \times \SU(2)_-$ R-charges $(j_+,j_-)$ and does not coincide with the BPS bound of D$(2,1|\upalpha)$ unless $j_+ = j_-$~\cite{deBoer:1999gea}. From worldsheet, supergravity, and integrability analyses~\cite{Eberhardt:2017fsi,Baggio:2017kza}, it is now understood that:
\begin{equation}
\label{eq:NSBPSbound}
    \quad \quad j_+ = j_- \, \, , \quad \textrm{(BPS states, NS sector)} \, ,
\end{equation}
where the $j_+ \neq j_-$ possibility is excluded in the NS sector. Furthermore, for certain fluxes, the dual CFT is argued to be a symmetric product orbifold~\cite{Eberhardt:2017pty,Gaberdiel:2018rqv,Eberhardt:2019niq,Gaberdiel:2024dva}. For general values of fluxes, following a suggestion of \cite{Gukov:2004ym}, Witten has recently provided more evidence~\cite{Witten:2024yod} that the CFT is a sigma model whose target space is the moduli space of instantons on $S^3 \times S^1$ by arguing it possesses a large $\mathcal{N}=4$ superconformal symmetry. 

In contrast to the NS sector, the R sector-- expected to include large extremal or near-extremal black holes-- remains poorly explored~\cite{Gukov:2004ym}; despite these sectors being related by spectral flow, the bulk physics at strong coupling appears very different and is worth exploring. Even for the BPS states, we know of no systematic, first-principled derivation (microscopic or otherwise) of the black hole entropy or the microscopic index in this sector. Our paper seeks to bridge this gap, in part using the results in~\cite{Heydeman:2025vcc}, by addressing three main goals:
\begin{itemize}
    \item Determine which charges $(j_+,j_-)$ permit BPS black holes (R sector BPS states) and compute their degeneracies at large charges. This can be achieved either using modular invariance of the dual CFT or using the bulk gravitational path integral. Special properties of the large $\mathcal{N}=4$ algebra prevent us from deriving this data using the spectral flow of \cite{Gukov:2004ym}. 
    \item Compute the spectrum of long $\mathcal{N}=4$ multiplets corresponding to non-BPS but near-extremal black holes, generalizing earlier studies~\cite{Heydeman:2020hhw,Boruch:2022tno,Heydeman:2024ezi} where there is a gap between the BPS and non-BPS multiplets, or \cite{Heydeman:2024fgk} where a tunable parameter causes the gap to close\footnote{Other models which may exhibit vanishing gaps and indices are supersymmetric SYK models \cite{Fu:2016vas,Heydeman:2022lse,Benini:2022bwa,Benini:2024cpf,Heydeman:2024ohc}. However, there is no known SYK model `dual' to a black hole which realizes small or large $\mathcal{N}=4$ supersymmetry in quantum mechanics. So we have no apriori expectations for the BPS degeneracies or non-BPS mass gap for any microscopic model preserving large $\mathcal{N}=4$ supersymmetry in the infrared.}. This information is not protected, hence there is no hope to derive our results by other means (other than solving the CFT dual exactly). 
    \item Construct complex black hole solutions which implement supersymmetric boundary conditions for any value of the temperature~\cite{Cabo-Bizet:2018ehj,Iliesiu:2021are,Chen:2023mbc,Cassani:2024kjn,Boruch:2025qdq}. This is particularly interesting for the $AdS_3 \times S^3 \times S^3 \times S^1$ background since the elliptic genus displays several unfamiliar features \cite{Gukov:2004fh}.
\end{itemize}

This work first recalls some of the classical features of black holes in $AdS_3 \times S_+^3 \times S_-^3 \times S^1$. The theory itself is labeled by the central charge $c$ and the parameter $\upalpha = Q_5^-/Q_5^+$, which crucially enters into the symmetry D$(2,1|\upalpha)$ as well as its affine extension $A_\gamma$. In purely classical supergravity $\upalpha$ can be continuous, but it is a rational number in string theory. The dependence on $\upalpha$ will ultimately be quite surprising once we consider quantum gravity effects. The BPS black hole solutions (with energy $E=J$, $J$ the angular momentum along $AdS_3$, and spins $(j_+, j_-)$ along $\SU(2)_+ \times \SU(2)_-$) have a classical entropy we denote by $S_0$. At small finite temperature, we can contrast the extremality bound with the BPS bound, finding
\begin{equation}
\label{eq:classicalBPSBHbound}
    \quad \quad j_- = \upalpha j_+ \, \, , \quad \textrm{(classical BPS black holes, R sector)} \, ,
\end{equation}
showing already that \eqref{eq:NSBPSbound} is too restrictive. This relation is analogous to the classical statement that only $J=0$ Reissner-Nordström black holes can be BPS, or the non-linear charge constraint for 1/16-BPS black holes in AdS$_5$. Beyond the classical approximation, the spectrum of near-BPS black holes is described by the large $\mathcal{N}=4$ Schwarzian theory solved in \cite{Heydeman:2025vcc}. The theory depends on a few parameters $\upalpha$, $S_0$ and $\Phi_r$ that can be extracted from the classical solution. $\Phi_r$ is the energy scale~\cite{Preskill:1991tb} at which the semiclassical Hawking calculation breaks down \cite{Iliesiu:2020qvm,Heydeman:2020hhw}; below this scale, black hole thermodynamics is modified, see \cite{Turiaci:2023wrh} for a pedagogical review. 

The Schwarzian partition function\footnote{There are two versions of the large $\mathcal{N}=4$ algebra. The one denoted $A_\gamma$ and used throughout this paper has \emph{linear} commutation relations, but a \emph{nonlinear} BPS bound. The other version $\tilde{A}_\gamma$ has a nonlinear algebra and BPS bound. The linearity of $A_\gamma$ is achieved by adding additional fermionic generators as well as an extra $U(1)$ to the nonlinear algebra. In this work we concern ourselves only with $A_\gamma$ because the higher dimensional realization of this theory necessarily has a linear algebra.} encodes predictions for the spectrum of Ramond-sector BPS and near-BPS states. The nonlinear BPS bound may be extracted from the algebra of conserved charges of the Schwarzian theory, and the nonlinearity allows multiple BPS $(j_+,j_-)$ superselection sectors with ground-state energies $E_{BPS}(j_+,j_-) \ne 0$. This is in stark contrast with small $\mathcal{N}=4$ cases where BPS states concentrate at $\SU(2)$ charge $j=0$ and $E=J$. The energy shifts which appear in the large $\mathcal{N}=4$ case are quantum gravity effects, despite referring to BPS states.

\smallskip

\noindent To summarize our central result, we find that the number of BPS multiplets labeled by spin $(j_+,j_-)$ in the Ramond sector\footnote{In our conventions the BPS supermultiplet labeled by $(j_+,j_-)$ \eqref{eq:XLN4SMA} has the largest spin states being $(j_+,j_-+1/2)$ and $(j_++1/2,j_-)$. See \cite{Petersen:1989pp} for a detailed description of the large $\mathcal{N}=4$ supermultiplet structure.} is given by:
\beq\label{eqn:NBPSAdS3intro}
N_{j_ + j_-} =\begin{cases}
    e^{S_0} \sin \left(\pi \frac{2 Q_5^- j_+ - 2 Q_5^+ j_- +Q_5^-}{Q_5^+ + Q_5^-} \right),~~~~\text{if}   ~~~(j_+,j_-)\in R_{\rm BPS},\\
     \hspace{.2cm} 0 \hspace{5.2cm}\text{if}   ~~~(j_+,j_-)\notin R_{\rm BPS}
    \end{cases}
\eeq
to leading order in the large charge limit, where
\beq
S_0 = 2\pi \sqrt{\frac{Q_1 Q_5^+ Q_5^-}{Q_5^+ +Q_5^-} J} \, ,
\eeq
is the naive Cardy entropy. We define the set
\beq
R_{\rm BPS} = \left\{ (j_+,j_-) \left| \, 0< \frac{2 Q_5^- j_+ - 2 Q_5^+ j_- +Q_5^-}{Q_5^+ + Q_5^-}<1\right.\right\}.
\label{eqn:BPSset}
\eeq
\begin{figure}
\centering
\begin{subfigure}{.5\textwidth}
  \centering
  \includegraphics[width=.73\linewidth]{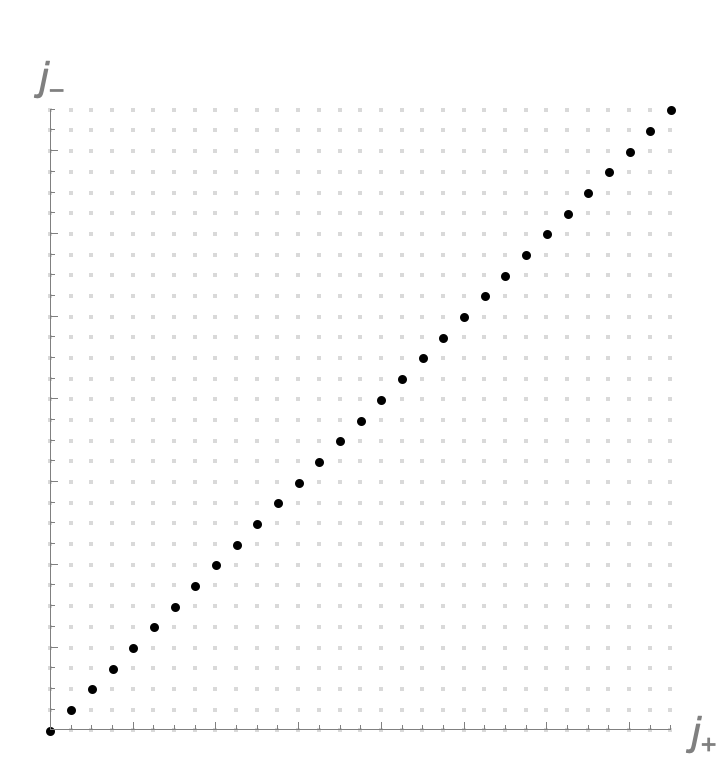}
  \caption{\footnotesize NS sector}
\end{subfigure}%
\begin{subfigure}{.5\textwidth}
  \centering
  \includegraphics[width=.73\linewidth]{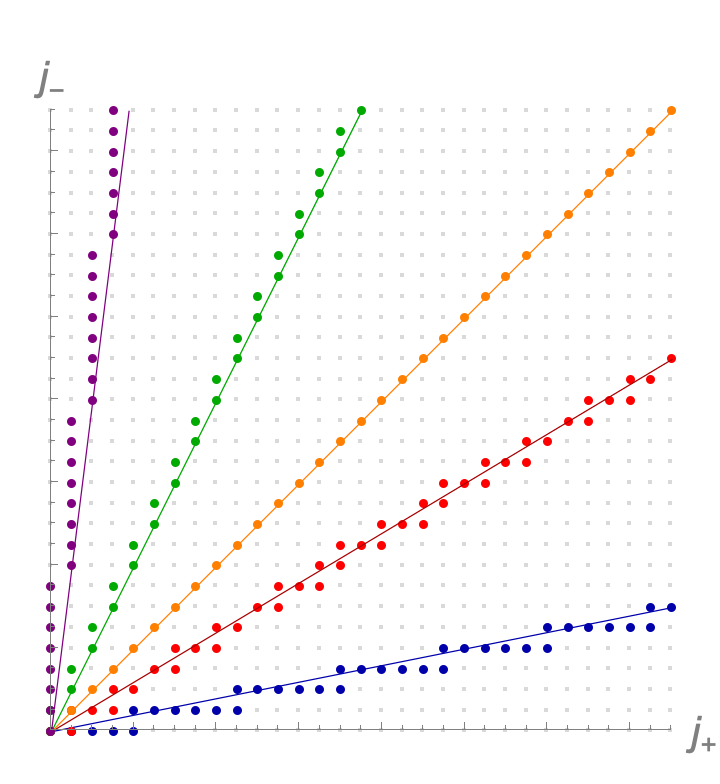}
  \caption{\footnotesize R sector}
\end{subfigure}
\caption{\footnotesize The BPS spectrum to the leading order in the semiclassical regime for a large $\cN=4$ theory is displayed. Each gray background point corresponds to a BPS multiplet labeled by spins $(j_+,j_-)\in \frac{1}{2}\mathbb{Z}_{\ge 0}\times \frac{1}{2}\mathbb{Z}_{\ge 0}$. \textbf{(a)} In the NS sector, the BPS multiplets are restricted to have $j_+=j_-$. \textbf{(b)} In the R sector, the allowed spins of BPS multiplets lie within the set defined in \eqref{eqn:BPSset}. Colored dots represent the BPS multiplets for different values of $ Q_5^-$ and $Q_5^+$, with increasing ratio from bottom to top: 
$Q_5^- / Q_5^+=0.2,\,0.6,\,1,\,2,\,8$, respectively; the straight lines, meanwhile, represent the expectations from classical gravity: $j_-\approx Q_5^- / Q_5^+ \,j_+$.}
\label{fig:BPSstates}
\end{figure}
The BPS degeneracy \eqref{eqn:NBPSAdS3intro} is always positive within this set. At large spin, the BPS condition asymptotes to $j_- \approx \upalpha j_+$, realizing the classical constraint \eqref{eq:classicalBPSBHbound} but remaining distinct from the NS sector result \eqref{eq:NSBPSbound}~\cite{Eberhardt:2017fsi}. As shown in Fig.~\ref{fig:BPSstates}, the BPS spectrum undergoes discrete jumps as the ratio of fluxes $\upalpha$ changes, with BPS states disappearing and reappearing with different spins. This structural novelty (which is completely absent in other examples) offers a sharp prediction for the dual CFT and motivates future investigation, especially since the full nature of the CFT is not yet understood. We also calculate the gap between the BPS states and the lightest non-BPS state in equation \eqref{eq:gapRLN4333}. This gap depends crucially on the value of $\upalpha$ and the total spin, and has an amalgam of features present with less supersymmetry~\cite{Stanford:2017thb,Heydeman:2020hhw,Boruch:2022tno,Turiaci:2023jfa,Johnson:2024tgg,Heydeman:2024ezi,Heydeman:2024fgk}.

There is one more interesting aspect of the ``large'' $\mathcal{N}=4$ Schwarzian theory which can also be understood from the higher-dimensional point of view. In general, the index or elliptic genus of a holographic CFT can be evaluated in gravity via a non-extremal BPS black hole \cite{Cabo-Bizet:2018ehj,Iliesiu:2021are,Cassani:2024kjn,Boruch:2025qdq}. A complex solution is typically necessary to implement supersymmetric boundary conditions for fermions in the presence of a Euclidean horizon. The solution depends on the temperature, but in general the on-shell action is temperature independent (or, more precisely, holomorphic in the case of the elliptic genus), matching the temperature independence of the Witten indices. In the present case, however, we construct non-extremal BPS black hole geometries in $AdS_3 \times S^3 \times S^3\times S^1$ that have \emph{non-holomorphic} on-shell actions\footnote{This can happen for black holes in nonsupersymmetric theories \cite{Chen:2023mbc}. An example of a temperature-dependent index is provided in \cite{Cabo-Bizet:2024gny} but so far it has not been given a clear gravity interpretation. }. The explanation is more obvious in the dual 2d CFT and is a consequence of the BPS bound depending non-linearly on the $R$-charges (the $E_{BPS}(j_+,j_-)$ introduces $\beta$ dependence in the partition function that is not removed by the $(-1)^{\sf F}$ insertion), but the result in gravity is surprising. In fact, we construct the Killing spinor and find it depends on both $\beta$ and $Q_5^\pm$. This is an example where knowledge of the extremal solution in gravity is not enough to capture all the information encoded in the elliptic genus, a fact overlooked in early treatments of these backgrounds  \cite{Gukov:2004fh,Gukov:2004ym}.

As a further extension of our results, we emphasize that the Schwarzian theory describes the dynamics of near-extremal horizons with an AdS$_2$ factor. An argument based on this theory is more general than one based on Virasoro symmetry, which requires an intermediate AdS$_3$ throat. For example, the small $\mathcal{N}=4$ Schwarzian analysis applies to black holes in $AdS_3 \times S^3 \times M$, with $M=K3$ or $T^4$, but also applies to the Reissner-Nordström black hole with no intermediate AdS$_3$ factor~\cite{Heydeman:2020hhw}. Similarly, we can comment on other large $\mathcal{N}=4$ backgrounds that present no AdS$_3$ factor. An example arising from intersections of branes in M-theory produces near-BPS black holes with near-horizon geometry $AdS_2 \times S^2 \times S^2 \times T^5$~\cite{Boonstra:1998yu}. We predict the black hole spectra by matching symmetries and parameters of the Schwarzian theory to this $AdS_2$ solution, despite the limited understanding of dual field theories in those cases.

\smallskip

\noindent The rest of the paper is organized as follows. Section~\ref{sec:STI} reviews the relevant Type IIB supergravity $AdS_3 \times S^3 \times S^3 \times S^1$ background, the large $\mathcal{N}=4$ supersymmetry, and aspects of the classical black hole solutions. In Section~\ref{sec:MAIN}, we introduce the large $\mathcal{N}=4$ Schwarzian theory and review its exact solution. Section~\ref{sec:BPSBHQC} presents our main results on the spectrum of BPS and near-BPS black holes. In Section~\ref{sec:NEBPSBH}, we construct non-extremal BPS black holes and show these solutions have the correct supersymmetric boundary conditions and reproduce the elliptic genus. We conclude in Section~\ref{sec:conc} with comments on the $AdS_2 \times S^2 \times S^2 \times T^5$ black hole and outline future directions. We also call attention to Appendix~\ref{app:N4Virasoro}, which is an alternate way of deriving many of these results using large $\mathcal{N}=4$ Virasoro symmetry, or equivalently the $AdS_3$ gravitational path integral.

\smallskip

\section{String Theory on $AdS_3 \times S^3 \times S^3 \times S^1$}\label{sec:STI}

In this section we review the $AdS_3 \times S^3 \times S^3 \times S^1$ background of Type II string theory. This case is famously subtle and shows new complications that are not present in the more familiar $AdS_3 \times S^3 \times M$ with $M=T^4$ or $K3$, in particular with respect to dual 2d CFT. From the bulk point of view, most recent literature focuses on fluctuations around the vacuum, but for our purposes we are more interested in the black hole solutions where supergravity is a reliable description and we can use Ramond fluxes in Type IIB. Before discussing quantum black holes in later sections, in this section we will present interesting classical aspects of black holes in $AdS_3 \times S^3 \times S^3 \times S^1$ that were not mentioned previously.

\subsection{The background and its asymptotic symmetries}

We follow the presentation of \cite{Gukov:2004ym} to review some features of the solution and set up notation. Rather than trying to take the decoupling limit of branes and fluxes, we will instead begin with a solution of Type IIB supergravity with a nontrivial metric and RR 3-form background (after the branes have backreacted). The relevant terms in the 10d action are
\beq\label{eq:typeIIBaction}
I = \frac{2\pi}{g_s^2}\int \sqrt{-g}\, e^{-2 \Phi} \,R - \pi \int F_3 \wedge * F_3 + \ldots,
\eeq
where $F_3 = \d C_2$ is the RR 3-form, $\Phi$ is the dilaton, and $g_s$ is the string coupling. We work in units where the string tension is $\alpha'=(2\pi)^{-2}$. The $AdS_3 \times S^3 \times S^3 \times S^1$ background metric is given by
\bea
\d s^2 &=& \ell^2\underbrace{ \frac{-\d t^2 + \d \varphi^2+ \d z^2}{z^2}}_{\text{unit $AdS_3$}}+ R_+^2 \underbrace{(\d \theta_+^2 + \sin^2 \theta_+ \d \phi_+^2 + \cos^2 \theta_+ \d \psi_+^2)}_{\text{unit $S_+^3$}} \nonumber\\
&&+ R_-^2 \underbrace{(\d \theta_-^2 + \sin^2 \theta_- \d \phi_-^2 + \cos^2 \theta_- \d \psi_-^2)}_{\text{unit $S_-^3$}}+ L^2 \underbrace{\d \theta^2}_{\text{unit $S^1$}},
\ea
where $0\leq \theta_\pm \leq \pi/2$ and $0\leq \phi_\pm,\psi_\pm \leq 2\pi$ and $\theta \sim \theta + 1$. The $AdS_3$ factor has $z>0$ and $\varphi\sim \varphi +2\pi$. The dilaton is constant for the solution we are interested in, and we normalize the string coupling so that $\Phi=0$. The 3-form is
\bea
F_3 &=& \lambda_0 \,\underbrace{z^{-3} \d t \wedge \d \varphi \wedge \d z}_{=\epsilon_0} \nonumber\\
&&+ \lambda_+ \, \underbrace{\frac{1}{4\pi^2}\sin 2\theta_+  \d \theta_+ \wedge \d \psi_+ \wedge \d \phi_+ }_{=\epsilon_+} \, \, +\,  \lambda_- \, \underbrace{\frac{1}{4\pi^2}\sin 2\theta_-   \d \theta_- \wedge \d \psi_- \wedge \d \phi_- }_{=\epsilon_-} ,
\ea
where $\epsilon_0$ is the volume form of a unit $AdS_3$ factor, and $\epsilon_\pm $ the volume form of the 3-spheres $S_\pm^3$ normalized so that $\int_{S_\pm^3} \epsilon_\pm = 1$. All other fields that were not written in \eqref{eq:typeIIBaction} vanish. 

The equation of motion for $F_3$ is automatically satisfied since $\nabla_\mu (F_3)_{\nu\rho\sigma}=0$. The equation of motion for the dilaton implies that the Ricci scalar of the metric vanishes. This constrains the radius of curvature of each factor:
\beq\label{eq:Ris0}
\frac{1}{6} R = - \frac{1}{\ell^2} + \frac{1}{R_+^2} + \frac{1}{R_-^2} = 0 \, .
\eeq
 The equations of motion arising from varying the metric are
$
R_{\mu\nu} = \frac{g_s^2}{4} (F_3)_{\mu\rho\sigma} (F_3)_{\nu}{}^{\rho \sigma},
$
and determine the coefficients that appear in the RR field $\lambda_0 = 4 \pi^2 \ell^2/g_s$, $\lambda_+ = 4 \pi^2 R_+^2/g_s $, and $\lambda_- = 4 \pi^2 R_-^2/g_s$.
Notice that this identification guarantees that the condition $R=0$ implies that $F_3 \wedge * F_3 = 0$, which is also a consequence of the dilaton equation of motion combined with the trace of Einstein's equation. 

We can identify the parameters above with the D1/D5-brane charges and fluxes. The flux through the two 3-spheres is
$$
Q_5^\pm = \int_{S_\pm^3} F_3 =  \frac{1}{g_s} \, 4 \pi^2 R_\pm^2,~~~~\Rightarrow~~~~ R_\pm = \frac{1}{2\pi} \sqrt{ g_s Q_5^\pm},
$$
where $Q_5^\pm$ are integers. This also fixes the value of $\lambda_\pm$ that appears in $F_3$ along the $S_\pm^3$ directions. \eqref{eq:Ris0} determines the $AdS_3$ radius of the solution
\beq
\ell = \frac{1}{2\pi} \sqrt{ \frac{g_s Q_5^+ \, g_s Q_5^-}{g_s Q_5^+ +g_s Q_5^- }},
\eeq
which also fixes $\lambda_0$ in terms of the charges and fluxes. The validity of the supergravity approximation requires a small string coupling but a large charge $g_s Q_5^\pm \gg 1$. Finally, the $S^1$ length $L$ can be written in terms of the $Q_1$ charge since
\beq
Q_1 = \int * F_3 = \frac{8 \pi^4 R_+^3 R_-^3 L}{\ell g_s} ,~~~~\Rightarrow~~~~ L =  \frac{4\pi\, g_s Q_1}{g_s Q_5^+ g_s Q_5^- \sqrt{ g_s Q_5^+ +g_s Q_5^-}}.
\eeq
Using these results, the 3-form can be written as
\beq
\label{eq:F3sol}
F_3 = Q_1 *\,(\epsilon_+ \wedge \epsilon_- \wedge \d \theta)+  Q_5^+\, \epsilon_+  + Q_5^- \, \epsilon_-.
\eeq
$Q_1$ becomes integral after incorporating quantum effects.  $g_s Q_1$, $g_s Q_5^+$ and $g_s Q_5^-$  fix the radius of the two 3-spheres and the length of the circle $S^1$, while the equations of motion determine the $AdS_3$ radius in terms of these parameters. In this regime, we see that the size of $S^1$ is small and the theory can effectively be described by its nine-dimensional compactification, unless $Q_1$ is much larger than the other charges.

\subsection*{Asymptotic symmetries of $AdS_3 \times S^3 \times S^3 \times S^1$}

To complete our review of the $AdS_3 \times S^3 \times S^3 \times S^1$ background we will describe its asymptotic symmetries. Large diffeomorphisms within the $AdS_3$ factor lead to a Virasoro algebra generated by the left-moving $L_n$ and right-moving $\bar{L}_n$, where $n\in \mathbb{Z}$ labels the Fourier momentum on the boundary circle with respect to
\beq
\label{eq:lightconecords}
w=\varphi-t\, ,~~~\text{and}~~~\bar{w}=\varphi +t \, ,
\eeq
respectively. To avoid a lengthy discussion of this well-known fact of asymptotically $AdS_3$ spaces, we simply mention that our conventions regarding the Virasoro algebra follow those in \cite{Kraus:2006wn}. The 3d Newton constant $G_N$ is given by
\beq
\frac{1}{16 \pi G_N} = \frac{2 \pi \text{Vol}(S_+^3 \times S_-^3 \times S^1)}{g_s^2},~~~~\text{Vol}(S_+^3 \times S_-^3 \times S^1)=( 2\pi^2 R_+^3)(2 \pi^2 R_-^3) L \, .
\eeq
The central charge of the Virasoro algebra, as determined by Brown and Henneaux, is given by
\beq
\label{eq:centralcharge}
c = \frac{3 \ell}{2G_N} = 6 Q_1 \frac{Q_5^+ Q_5^-}{Q_5^+ + Q_5^-} \, .
\eeq

The compact space $S^3 \times S^3 \times S^1$ and its isometries give rise to other asymptotic symmetries related to rotations along those directions. From the point of view of $AdS_3$, they are realized as large gauge transformations of Kaluza-Klein gauge fields. These gauge fields can be decomposed into a flat component, described by Chern-Simons theory on $AdS_3$, and the non-flat remainder. For the discussion on asymptotic symmetries, only the Chern-Simons component is relevant and leads to a boundary Kac-Moody algebra. The level of these boundary symmetries can be extracted from the coefficient of the Chern-Simons action. The results were given in \cite{Gukov:2004ym} without derivation, and we provide a simple derivation using the approach of Hansen and Kraus \cite{Hansen:2006wu} in Appendix \ref{app:SU2DER}.  

The isometries of $S^3 \times S^3$ give rise to the group $\SO(4) \times \SO(4)$, or equivalently to four $\SU(2)$ gauge fields. It is useful to parametrize the two three-spheres by coordinates $y_\pm^i$, with $i=1,\ldots,4$ such that $y_\pm^i y_\pm^i=1$. For either sphere, they are related to our previous coordinates via $y^1 = \sin \theta \sin \phi$, $y^2=\sin \theta \cos \phi$, $y^3=\cos \theta \sin \psi$, $y^4=\cos \theta \cos \psi$. The metric with the KK gauge field turned on is
\beq
\d s^2 = \d s^2_{AdS_3} + \sum_{\pm} R_\pm^2(\d y_\pm^i - A_\pm^{ij} y_\pm^j)(\d y_\pm^i - A_\pm^{ik}y_\pm^k)  + L^2 \d \theta^2,
\eeq
where $A^{ij}_\pm$ are one-forms along the asymptotically $AdS_3$ spacetime, and transform in the adjoint of the two $\SO(4)$ isometries for each three-sphere. It is convenient to write the result in terms of two copies of $\SU(2)_L \times \SU(2)_R$. For a given $\SO(4)$ of $S^3$ gauge field, the two basis are related by
\beq\label{eq:SOTOSU}
A^{I4} = -\frac{1}{2} (A_L^I-A_R^I),~~~A^{IJ}= - \frac{1}{2} \epsilon^{IJK}(A_L^K+A_R^K),~~~~I,J,K=1,2,3 \, ,
\eeq
where we removed the $\pm$ indices to avoid clutter. In addition to this modification of the metric, it is also necessary to modify $F_3$ in the presence of the $\SU(2)$ gauge fields. The precise form that $F_3$ takes in terms of $A_\pm^{ij}$ is crucial for the derivation of the action of these modes and can be found in the appendix \ref{app:SU2DER}. In that appendix we show that the action of these $\SU(2)$ modes is
\beq
I \supset  -\i k^+  \int_{AdS_3} {\rm CS}(A_L^+)+\i k^+  \int_{AdS_3} {\rm CS}(A_R^+) + (+\to -),
\eeq
where ${\rm CS}(A) =\frac{1}{4\pi} \operatorname{Tr}(A \d A + \frac{2}{3} A^3)$ is the Chern-Simons form, and the levels are
\beq
k^+ = Q_1 Q_5^+ ,~~~~~k^- = Q_1 Q_5^-.
\eeq
To derive the boundary symmetry algebra, we need to specify boundary conditions, and we again follow  \cite{Kraus:2006wn}. Let us focus on the $\SU(2)_L^\pm$ gauge fields. If we work in the grand canonical ensemble, we fix the components $A_{L\bar{w}}^{\pm I}$ at the $AdS_3$ boundary. Without loss of generality, we align the axis of rotation such that the only nonzero component is
\beq
A_{L}^{\pm 3}|_{\rm bdy} = \Omega_L^\pm \, \d\bar{w} + \ldots, 
\eeq
where $\Omega_L^\pm$ are the two $\SU(2)^\pm$ angular velocities. The boundary terms necessary to implement this are described in \cite{Kraus:2006wn,Hansen:2006wu}. Large gauge transformations consistent with this boundary condition give rise to a pair of left-moving $\widehat{\SU}(2)_{k^\pm}$ Kac-Moody symmetries whose generators we denote by $T^{\pm I}_n$ or in bi-spinor notation $T^{AB}_n$ and $T^{\dot{A}\dot{B}}_n$. These generators are determined by the Fourier modes of the second boundary component $A_{L w}^{\pm I}$. The same considerations apply to $\SU(2)_R^+ $ and $\SU(2)_R^-$. The appropriate boundary conditions now fix $A_{Rw}^{\pm I}$ and therefore give rise to the right-moving generators we denote by $\bar{T}_n^{\pm I}$. 

There are also large supersymmetry transformations generated by left- and right-moving spin-3/2 currents we denote by $G^{A\dot{B}}_n$ and $\bar{G}^{A\dot{B}}_n$ with $n\in \mathbb{Z}$ or $\mathbb{Z}+1/2$, depending on whether we work in the R or NS sector, respectively \cite{deBoer:1999gea}. The left-moving generators are in the bifundamental of $\SU(2)_L^+ \times \SU(2)_L^-$ while the right-moving generators are in the bifundamental of $\SU(2)_R^+ \times \SU(2)_R^-$. Their algebra is given 
\bea
\label{eq:commutatorsGG}
\{ G_n^{A \dot{B}}, G_m^{C\dot{D}}\} &=& -\frac{1}{2} \epsilon^{\dot{B} \dot{D}} \epsilon^{AC} \left[ 2 L_{n+m} + \frac{c}{3} \delta_{n+m} \Big(m^2-\frac{1}{4}\Big)\right],\nonumber\\
&&+(n-m) \left[  \frac{k^+}{k^++k^-} T^{\dot{B} \dot{D}}_{n+m} \varepsilon^{AC} + \frac{k^-}{k^++k^-} T^{AC}_{n+m} \varepsilon^{\dot{B}\dot{D}}\right].\label{eq:GGVIRASORO}
\ea
and similarly for the right movers.

There are also $\U(1)$ symmetries analyzed in detail in Section 9 of \cite{Gukov:2004ym}. One arises from the metric $\d s^2 \supset L^2 (\d \theta + a)^2$ with $a$ a one-form along $AdS_3$, and another one-form $b$ arises from the RR field $F_3 \supset \d b \wedge \d \theta$. They lead to two $\U(1)$ Kac-Moody algebras, with generators we denote $U_n$ for their left-moving component and $\bar{U}_n$ for the right-moving one, which are related to linear combinations of large gauge transformations of $a$ and $b$. The level $k_{\U(1)}$ of the $\U(1)$ symmetry obtained from the Chern-Simons term in gravity is $k_{\U(1)}=k^+ + k^-$.

The $\U(1)$ bosonic generators come with spin-1/2 fermionic partners commonly denoted $Q^{A\dot{A}}_n$ and $\bar{Q}^{A\dot{A}}_n$ in the bifundamental of $\SU(2)_L^+ \times \SU(2)_L^-$ and $\SU(2)_R^+ \times \SU(2)_R^-$ respectively. It is conventional to take them anti-hermitian \cite{Petersen:1989pp,Petersen:1989zz}. These are the partners of the $\U(1)$ modes under the fermionic $G$-supersymmetry, namely
\beq
\label{eq:commutatorsUGQ}
[U_m, G_n^{A\dot{A}} ] = \i \frac{m}{2} Q^{A\dot{A}}_{m+n} \, ,~~~\text{and}~~~\{ Q^{A\dot{B}}_n, Q^{C\dot{D}}_m\} = (k^++k^-)\varepsilon^{\dot{B}\dot{D}}\varepsilon^{AC}\delta_{n+m,0} \, .
\eeq
The commutation relations with $T^\pm$ are the standard ones based on their representation, but they do affect some commutation relations we did not write yet, such as
\beq
[ T^{+,I}_m , G^{A\dot{B}}_n] = \frac{1}{2} (\sigma^I)_C^{~A}\Big(G^{C\dot{B}}_{m+n} +\frac{k^+}{k^++k^-} m Q^{C\dot{B}}_{m+n} \Big) \, ,\label{eq:TGVIRASORO}
\eeq
and similarly for the other $\SU(2)$ generator. Finally, there is one more anticommutator we need to specify:
\beq\label{eq:QGVIRASORO}
\{Q^{A\dot{B}}_n,G^{C\dot{D}}_m\}= T_{n+m}^{AC} \varepsilon^{\dot{B}\dot{D}} -  T_{n+m}^{\dot{B}\dot{D}} \varepsilon^{AC} +  \i \varepsilon^{AC} \varepsilon^{\dot{B}\dot{D}} U_{n+m} \, .
\eeq
There are the main non-trivial commutation relations between the generators, and we will use them later to match the results from the Schwarzian theory.

The complete superconformal algebra (we focus on left-movers and similar considerations apply to right-movers) is denoted $A_\gamma$ and includes the generators
\begin{align}
\label{eq:Agammagenerators}
A_\gamma:~~~~~~~L_n \, ,\quad ~G^{A\dot{A}}_n,\quad~T^{AB}_n,\quad~T^{\dot{A}\dot{B}}_n,\quad~ Q^{A\dot{A}}_n,\quad ~U_n\, ,
\end{align}
The generators in the NS sector include a global subgroup $\text{D}(2,1|\upalpha)$ and $\gamma=\upalpha/(1+\upalpha)$. This subgroup of $A_\gamma$ annihilates the vacuum $AdS_3 \times S^3 \times S^3 \times S^1$ solution. The existence of the $\text{D}(2,1|\upalpha)$ global symmetry can already be inferred just from the Killing vectors and spinors of the empty AdS solution~\cite{Gauntlett:1998kc}. As we explained above, the affine extension comes from the boundary gravitons~\cite{Brown:1986nw}, so we have Virasoro symmetry as well as a pair of $\SU(2)$ current algebras. The central charge and $\upalpha$ are given in terms of the $\SU(2)$ levels by the relations
\beq
c = \frac{6 k^+ k^-}{k^+ + k^-} \, ,~~~~~\upalpha = \frac{k^-}{k^+}= \frac{Q_5^-}{Q_5^+} = \frac{R_-^2}{R_+^2} \, .
\eeq
The relationship between these parameters is a consequence of the fact that the fermionic asymptotic symmetries combine with bosonic generators to make the large $\mathcal{N}=4$ algebra. The parameter $\upalpha$ can be interpreted as the ratio between the size of the two $S^3$.   

The additional $U$, $Q^{A\dot{A}}$ in \eqref{eq:Agammagenerators} beyond the naive set of super-Virasoro generators allow us to realize the algebra linearly (rather than non-linearly \cite{Knizhnik:1986wc,Schoutens:1988ig,Schoutens:1988tg}). Therefore, the algebra can be decomposed as $A_\gamma = \tilde{A}_\gamma \times A_{\mathcal{S}}$ where $A_{\mathcal{S}}$ is generated by $Q^{A\dot{A}}$ and $U$. The generators written in this decomposition are not exactly the same as those that appear in $A_\gamma$; the precise map can be found e.g. in equation (3.5) of \cite{deBoer:1999gea}. In removing the $A_\mathcal{S}$ factor the central charge and levels of the remaining generators become
$
\tilde{c} = c-3,$
 and 
$\tilde{k}^\pm = k^\pm -1.
$
This will be important in Appendix \ref{app:N4Virasoro} when we compute the one-loop gravitational path integral over $AdS_3$, but in the near-extremal limit, relevant for the Schwarzian theory, we will take the large level limit and these shifts can be ignored.

\subsection{The classical black hole geometry}\label{sec:10dbh}
Having described the vacuum and asymptotic symmetries, we now consider rotating black hole solutions in $AdS_3 \times S^3 \times S^3 \times S^1$. We consider boundary conditions on the spatial circle in $AdS_3$ to be in the RR sector for simplicity, which is possible because this circle is not contractible in the black hole background. We also focus on the case with a vanishing $\U(1)$ charge, which we denote by $u$ and $\bar{u}$ for left and right mover components respectively. We work in an ensemble with $u=\bar{u}=0$ but the generalization to $u,\bar{u}\neq 0$ is straightforward. 

The (Lorentzian) geometry which solves the equations of motion of \eqref{eq:typeIIBaction} is
\bea
\label{eq:BHmetricALR}
\d s^2 &=& - f \d t^2 + \frac{\ell^2 \d r^2}{f} + r^2 \Big[\d \varphi  - \frac{r_- r_+ }{r^2} \d t\Big]^2 \nonumber\\
&&+R_+^2 \Big[\d \theta_+^2 + \sin^2 \theta_+ \Big(\d \phi_+ +  \frac{A_L^{+\,3}+A_R^{+\,3}}{2}\Big)^2 +\cos^2 \theta_+ \Big(\d \psi_+ + \frac{A_L^{+\,3}-A_R^{+\,3}}{2}\Big)^2\Big] \nonumber\\
&& + R_-^2 \Big[\d \theta_-^2 + \sin^2 \theta_- \Big(\d \phi_- +  \frac{A_L^{-\,3}+A_R^{-\,3}}{2}\Big)^2 +\cos^2 \theta_- \Big(\d \psi_- + \frac{A_L^{-\,3}-A_R^{-\,3}}{2}\Big)^2\Big]\nonumber\\
&& + \, L^2 \d \theta^2 \, , \label{eq:BHMET10d}
\ea
where $f=(r^2-r_+^2)(r^2-r_-^2)/r^2$ and the four $\SU(2)$ gauge fields are given by
\beq
A_L^{\pm 3} = 2 \Omega^\pm_L\,   \frac{\d \varphi - \Omega \d t}{1-\Omega},~~~~~A_R^{\pm 3} = 2 \Omega^\pm_R  \, \frac{\d \varphi - \Omega \d t}{1+\Omega}.
\eeq
We have used the $\SO(4)_+ \times \SO(4)_-$ symmetry to simplify the form of the solution without loss of generality. The profile of $F_3$ takes the form \eqref{eq:F3sol}, but the volume forms on the spheres are evaluated on the new metric that includes the KK gauge fields. 

The solution depends on a few parameters whose physical interpretation is easy to identify. The first factor of \eqref{eq:BHmetricALR} becomes asymptotically $AdS_3$ as $r \to\infty$, while the Kaluza-Klein gauge fields become $A_L^{\pm3} = \Omega_L^\pm \d\bar{w} + \ldots$ and $A_R^{\pm 3} = \Omega_R^\pm \d w + \ldots$, consistent with the boundary conditions specified in the previous section. This identifies $\Omega_{L/R}^\pm$ with angular velocities on $S^3 \times S^3$. The event horizon is located at $r=r_+$ and the  contractible circle is identified via
$
(t,\varphi) \sim (t-\i \beta, \varphi-\i \beta \Omega),
$
where 
\beq
\beta=\frac{2\pi \ell r_+}{r_+^2-r_-^2},~~~\text{and}~~~\Omega=r_-/r_+,
\eeq
with all other $S^3 \times S^3 \times S^1$ angles kept fixed. $\varphi \sim \varphi + 2\pi$ is non-contractible and compact in our solution. The horizon is generated by the Killing vector $v= \partial_t + \Omega \partial_\varphi$, and the metric of the two 3-spheres is smooth since they satisfy $i_v \, A_{L/R}^\pm = 0$ at the horizon. 

Following the conventions of \cite{Hansen:2006wu}, the observations in the previous paragraph imply that the rotating black hole contributes to the boundary partition function
\beq
\label{eq:ZTrgravityvars}
Z={\rm Tr} \,\left[ e^{-\beta H - \beta \Omega J} \, e^{\beta \Omega^+_L J_L^{+3}}e^{\beta \Omega^-_L J_L^{-3}} \, e^{\beta \Omega^+_R J_R^{+3}}e^{\beta \Omega^-_R J_R^{-3}} \right],
\eeq
in the RR sector, where we turned off the $\U(1)$ charges with generator $U$ in \eqref{eq:Agammagenerators}, but this analysis can be easily generalized. In this expression, we introduce the boundary Hamiltonian $H=L_0 +\bar{L}_0 - c/12$, angular momentum $J=\bar{L}_0-L_0$, and $J_{L}^{\pm a} = T_0^{\pm a}$ and $J_R^{\pm a} = \bar{T}_0^{\pm a}$.  This black hole not only carries angular velocity, but, due to the Chern-Simons terms that appear in the action, also carries angular momentum along the compact space. 

The holographic dual to the asymptotically $AdS_3 \times S^3 \times S^3 \times S^1$ geometry is a two-dimensional CFT with large $\mathcal{N}=4$ supersymmetry. In this context, it is useful to introduce the modular parameters of the boundary torus
\beq
\label{eq:leftrighttemperatures}
\tau = \frac{\i\beta(1-\Omega)}{2\pi}=\frac{\i \beta_L}{2\pi}, \quad ~q=e^{2\pi \i \tau},~~~{\rm and}~~~\bar{\tau}=-\frac{\i\beta(1+\Omega)}{2\pi}=\frac{\i \beta_R}{2\pi}, \quad ~\bar{q}=e^{-2\pi \i \bar{\tau}}.
\eeq
$\beta_L$ and $\beta_R$ are the left- and right-moving inverse temperatures, respectively. We define the rescaled chemical potentials $\alpha_-$ and $\alpha_+$ (and their right-moving counterparts) for convenience to be
\beq
4\pi \i \alpha_\pm = \beta \Omega_L^\pm ,~z_\pm = e^{4\pi \i \alpha_\pm},~~~\text{and}~~~4\pi \i\bar{\alpha}_\pm = \beta \Omega_R^\pm,~~\bar{z}_\pm = e^{-4\pi \i \bar{\alpha}_\pm}. \label{eq:OmegaAlphasss}
\eeq
For solutions that are real in Euclidean signature, the barred and unbarred quantities are complex conjugate to each other and all chemical potentials are imaginary. For more general reality conditions, they are independent variables. In terms of these parameters, the partition function \eqref{eq:ZTrgravityvars} introduced above can be written as
\beq
Z={\rm Tr} \,\left[ q^{L_0 -\frac{c}{24}} \, z_+^{2 T^{+3}_{0}}\, z_-^{2 T^{-3}_{0}}\, \bar{q}^{\bar{L}_0-\frac{c}{24}}\, \bar{z}_+^{2 \bar{T}^{+3}_{0}}\, \bar{z}_-^{2 \bar{T}^{-3}_{0}}\right].
\eeq
Finally, we also denote by $h$ and $\bar{h}$ the eigenvalues of $L_0$ and $\bar{L}_0$ and they determine the mass and spin along $AdS_3$ of the black holes.

The on-shell action for this black hole as a function of temperature and chemical potentials gives a prediction for the grand canonical partition function of the boundary 2d CFT in the large $c$ limit. This can be done directly at the 10d level, but it is more illuminating to first recast the theory on its $AdS_3$ factor, compactifying on $S^3 \times S^3\times S^1$, since the evaluation of the action of 3d gravity coupled to Chern-Simons fields is straightforward \cite{Kraus:2006wn}. The result for the action is
\beq
\label{eq:fullonshellaction}
\log Z \sim - I_{\rm on-shell} = \frac{\i \pi c}{12 \tau} - \frac{2\pi \i k^+}{\tau} \alpha_+^2 - \frac{2\pi \i k^-}{\tau} \alpha_-^2 + \text{h.c.}
\eeq
The first term arises from the metric, while the second and third arise from the Chern-Simons action. This free energy can be used to extract the ADM charges of the solution if necessary. 

From the point of view of the effective 3d gravity theory, most of the spin/charge dependence of the solution may naively be removed by a gauge transformation. However, thanks to the Chern-Simons terms present in the action, there are still nontrivial charges carried by the black hole. It is particularly instructive to evaluate $L_0$ and $\bar{L}_0$ in terms of the temperature and the $S^3\times S^3$ angular momenta. We obtain for vanishing $\U(1)$ charge
\beq
h-\frac{c}{24} = \frac{j_-^2}{k^-}+\frac{j_+^2}{k^+} + \frac{c \pi}{6 \beta_L^2} \, ,
\eeq
and similarly for $\bar{L}_0$. At extremality, we have by definition $r_- \rightarrow r_+$, so that in \eqref{eq:leftrighttemperatures} we have $\beta_L\to\infty$. Therefore, the condition that the black hole is extremal is the condition:
\beq
h^{ext}-\frac{c}{24} =\frac{j_-^2}{k^-}+\frac{j_+^2}{k^+} \, .
\eeq

On the other hand, the BPS bound determined by preserving supersymmetry (again in the case that $U=0$) at the classical level is
\beq
h^{BPS} -\frac{c}{24} = \frac{(j_++j_-)^2}{k^++k^-} \, ,~~~~~\text{(classical)}.
\eeq
This can be derived either from an analysis of the Killing spinors, or as a limit of the quantum BPS bound we review in the next section. When we approach the extremal BPS limit the geometry develops a long $(AdS_2 \times S^1) \times S^3 \times S^3 \times S^1$ throat (up to mixing between factors due to rotation) with an emergent D$(2,1|\upalpha)$ isometry. Notice that this is not a subset of the boundary Virasoro generators since the black hole lives in the Ramond sector while D$(2,1|\upalpha)$ is a subset of $A_\gamma$ only in the NS sector. Nonetheless, D$(2,1|\upalpha)$ does appear as an approximate isometry near the horizon and plays an important role in the next section.

The extremal and BPS $L_0$ bounds at the classical level only match when
\beq
j_- = \frac{k^-}{k^+} j_+ \, ,~~~~~\text{(classical)}.\label{eq:CLASSJJBPS}
\eeq
What is the meaning of this relation? In general, for causally well behaved solutions in Lorentzian signature, all supersymmetric black holes are extremal, but not all extremal black holes are supersymmetric. When the black hole has several charges, the intersection of the supersymmetric and extremal loci in parameter space determines which set of microscopic charges are compatible with a BPS black hole solution. Therefore, \eqref{eq:CLASSJJBPS} is the $AdS_3 \times S^3 \times S^3 \times S^1$ analog of the fact that BPS black holes in 4d flat space have vanishing angular momentum, or the non-linear constraint for $AdS_5$ black holes \cite{Chong:2005hr,Cabo-Bizet:2018ehj}. Actually, if one allows for genuine complex solutions, one can indeed find supersymmetric solutions which are not extremal; these give an alternative way to compute the index, and we undertake this in Section \ref{sec:NEBPSBH}.

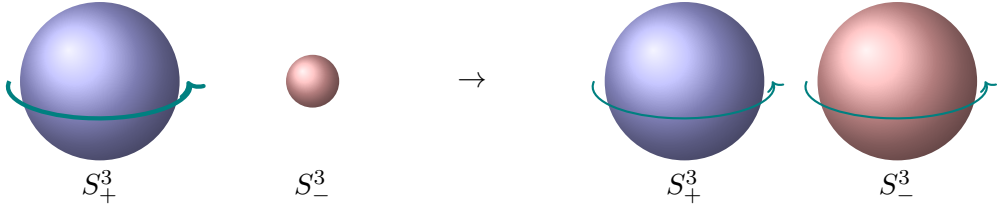
\begin{figure}[t!]
    \centering
\begin{tikzpicture}[scale=0.7, every node/.style={scale=1}]
    \shade[ball color=blue!30] (0,0) circle (1.5cm);
    \node at (0,-2) {$S_+^3$};
    \shade[ball color=red!30] (4,0) circle (0.5cm);
    \node at (4,-2) {$S_-^3$};    
    \draw[->, ultra thick, teal] (-1.7,0) arc[start angle=170, end angle=370, x radius=1.7cm, y radius=.6cm];
    \node at (7,0) {$\rightarrow$};    
\end{tikzpicture}
~~~~~~~
\begin{tikzpicture}[scale=0.7, every node/.style={scale=1}]
    \shade[ball color=blue!30] (0,0) circle (1.5cm);
    \node at (0,-2) {$S_+^3$};
    \shade[ball color=red!30] (4,0) circle (1.5cm);
    \node at (4,-2) {$S_-^3$};    
    \draw[->, thick, teal] (-1.7,0) arc[start angle=170, end angle=370, x radius=1.7cm, y radius=.6cm];
    \draw[->, thick, teal] (4-1.7,0) arc[start angle=170, end angle=370, x radius=1.7cm, y radius=.6cm];
\end{tikzpicture}
    \caption{ \textit{Left:} Case with $R_- \ll R_+$, or equivalently $Q_5^- \ll Q_5^+$. To leading order in $G_N$, BPS states rotate mostly on the larger sphere $S_+^3$. \textit{Right:} Case with comparable $R_+ \sim R_-$ or equivalently $Q_5^- \sim Q_5^+$. To leading order, BPS states distribute their angular momentum mostly equally.  }
    \label{fig:D1D5D5_BPS states}
\end{figure}

In figure \ref{fig:D1D5D5_BPS states} we show how the spin depends on the size of the 3-spheres for the BPS black holes. The relation \eqref{eq:CLASSJJBPS} is incompatible with spin quantization, but similarly to the $AdS_5$ case, we should remember that this relation is only valid in the semiclassical $j \sim \mathcal{O}(c)$ limit. A major result of this paper is that we find a refinement of this formula valid for order one spins; to do this we will need to describe the quantum gravity corrections that take place near extremality and compute the quantum corrections to the near-BPS black hole spectrum. 

We will need one more piece of information. Given fixed angular velocities, the solution written above is not the most general configuration given the choice of boundary conditions. This can be seen in multiple ways. On one hand, since the angular momentum should be half-integer, the partition function can only depend on $\beta \Omega_{L/R}^\pm$ mod $4\pi \i \mathbb{Z}$. Therefore, all saddles related by $\beta\Omega_{L/R}^\pm \to \beta\Omega_{L/R}^\pm + 4\pi \i \mathbb{Z}$ should be summed over, since they are physically distinct in the bulk. This corresponds to integer spectral flow on the 2d CFT side. Another way to see this is that the boundary condition for $A_{L/R}^\pm$ specifies the holonomy around a boundary circle; not the connection itself, but its exponentiated form. The exponentiation appearing here allows for the same integer shifts of the angular velocities as dictated by the half-integrality of spin. In terms of $\alpha_\pm$ and $\bar{\alpha}_\pm$ these shifts act as $\alpha_+ \to \alpha_+ +\mathbb{Z}$ and similarly for the others; see Appendix \ref{app:N4Virasoro}. One can think of these geometries as being subdominant saddle points which are required for charge quantization~\cite{Aharony:2021zkr}.

\section{Spectrum of BPS black holes in $AdS_3 \times S^3 \times S^3 \times S^1$}\label{sec:MAIN}

So far we have presented the classical analysis of black holes in
$AdS_3 \times S^3 \times S^3 \times S^1$ and considered their BPS regime. We can refine this analysis by incorporating quantum gravity corrections, which become particularly relevant near extremality both for BPS and near-BPS black holes. As the system approaches the extremal limit, the black hole throat becomes infinitely long and develops a nearly $AdS_2$ region. It signals the emergence of 1D superconformal symmetry and suggests that the low-energy dynamics and the quantum corrections in this regime are captured by the corresponding Schwarzian theory~\cite{Nayak:2018qej,Moitra:2018jqs,Castro:2018ffi,Sachdev:2019bjn,Ghosh:2019rcj,Iliesiu:2020qvm,Heydeman:2020hhw,Boruch:2022tno,Iliesiu:2022onk}.

Here, we begin in section \ref{sec:LargeN4Review} with a review of the large $\mathcal{N}=4$ super-Schwarzian theory, introduced in \cite{Heydeman:2025vcc}, which describes the dynamics of near-BPS black holes with emergent near-horizon isometry supergroup D$(2,1|\upalpha)$. This has a bosonic subgroup containing a factor of the $\SL(2,\mathbb{R})$ conformal group (acting on AdS$_2$) and $\SU(2)_+ \times \SU(2)_-$ acting as the $R$-symmetry. The fermionic part is generated by four fermions in the bi-fundamental representations of the two $\SU(2)$ factors. Crucially, compared to simpler examples, the algebra (and thus the Schwarzian theory) depends on a continuous parameter $\upalpha$, which is always a rational number in the string theory examples. We outline the construction of the super-Schwarzian and describe its spectrum, leaving more details in~\cite{Heydeman:2025vcc}. This discussion of the Schwarzian is more general than the specific application we have in mind, which are the near extremal black holes in $AdS_3 \times S^3 \times S^3 \times S^1$. The application to that background is discussed in section~\ref{sec:BPSBHQC}; where we will see a number of surprising features for both the BPS and non-BPS black holes that are not present classically.

\subsection{The Large $\mathcal{N}=4$ Schwarzian Theory}\label{sec:LargeN4Review}

The large $\mathcal{N}=4$ super-Schwarzian admits two constructions, corresponding to the two possible versions of the large $\mathcal{N}=4$ Virasoro algebra (see the discussion below \eqref{eq:Agammagenerators}). The ``fields'' of the Schwarzian theory are super-reparametrizations labeled by this choice of algebra (in the NS sector from a 2d CFT point of view), whereas the conserved charges may be extracted by the Noether procedure from the action (and are essentially in the R sector from the 2d CFT point of view).

In the minimal Schwarzian theory based on $\tilde{A}_\gamma$, the conserved charges satisfy a non-linear algebra. In the non-minimal $A_\gamma$ theory, obtained by adding a decoupled $\U(1)$ mode together with a set of fermions in the bi-fundamental of $\SU(2)$ as in \eqref{eq:Agammagenerators}, there is a linear algebra of conserved charges. As we discussed in \cite{Heydeman:2025vcc}, a linear algebra of conserved charges is more appropriate for the gravitational setup from the higher-dimensional perspective, and indeed this is true for the  example of $AdS_3 \times S^3 \times S^3 \times S^1$. Therefore, in the following, we describe the action and spectrum of the linear theory only. 

We can construct the action of the theory either from a purely quantum-mechanical perspective \cite{Kozyrev:2021agn,Kozyrev:2021icm}, or via the BF formulation of JT gravity with super-isometry $\text{D}(2,1|\upalpha)$ and the addition of the $\U(1)$ multiplet \cite{Heydeman:2025vcc,Heydeman:2020hhw}. In either case, the bosonic sector of the theory is 
\bea\label{eq:actionbosonicint}
I &=& S_0 - \Phi_r \int \d\tau \,\Big\{\, \text{Sch}(f,\tau)+ \frac{1}{\gamma} \, {\rm Tr} [ (g_+^{-1} \partial_\tau g_+)^2]+\frac{1}{1-\gamma} \, {\rm Tr} [ (g_-^{-1} \partial_\tau g_-)^2] \nonumber\\
&&\hspace{3cm}\left.  +\frac{1}{2\gamma(1-\gamma)} (g_0^{-1}\partial_\tau g_0)^2 \right\} + \text{(fermions)} \, ,
\ea
where following standard practice we introduce the parameter $\gamma$ via 
\begin{align}
\label{eq:alphagammadef}
   \upalpha=(1-\gamma)/\gamma \, . 
\end{align} 
$\text{Sch}(f,\tau)$ denotes the Schwarzian derivative of the reparametrization $f(\tau)$. $g_\pm(\tau)\in \SU(2)_\pm$ is the degree of freedom associated with the breaking of local $\SU(2)_\pm$ transformations, and $g_0(\tau) \in \U(1)$ is the decoupled $\U(1)$ mode. The relative values of the couplings in each term are fixed by supersymmetry, while the overall normalization $\Phi_r$ is a free parameter determined by matching to the higher-dimensional black hole solution (which we compute in the next section). The fermionic completion of the above action has a partner of the reparametrization mode $\eta^{A\dot{A}}(\tau)$ where $A$ ($\dot{A}$) is an index in fundamental of $\SU(2)_+$ ($\SU(2)_-$). The partners of the $\U(1)$ mode are also in the bi-fundamental $\psi^{A\dot{A}}(\tau)$ and are decoupled from all other fields. The explicit form of the full action was worked out in \cite{Kozyrev:2021icm}.

We are interested in the quantization of the large $\mathcal{N}=4$ Schwarzian theory. The partition function has a Hilbert space interpretation 
\beq
Z(\beta,\alpha_+,\alpha_-,\mu) = {\rm Tr} \left[ e^{-\beta H} e^{4\pi \i \alpha_+ J^3_+} e^{ 4\pi \i \alpha_- J^3_-} e^{2\pi \i \mu U} \right].
\eeq
$H$ is the Hamiltonian, $J_\pm^I$ are the generators of $\SU(2)_\pm$, and $U$ is the $\U(1)$ generator. In the grand canonical ensemble, $\beta$ is the inverse temperature and $\alpha_\pm$ and $\mu$ are therefore proportional to chemical potentials associated with the bosonic symmetries. In \cite{Heydeman:2025vcc}, we obtained the result:
\beq
Z = \sum_{n,m \, \in \, \mathbb{Z}} \, \, \sum_{r \,\in \, \mathbb{Z}/u_0} Z_{\text{1-loop}} \, e^{{S_0+\frac{2\pi^2\Phi_r}{\beta}\left(1- \frac{4(1+\upalpha)}{\upalpha}(\alpha_+ +n)^2-4(1+\upalpha)(\alpha_- +m)^2-\frac{(1+\upalpha)^2}{\upalpha}(\mu+r)^2\right)}} \, .\label{eqn:LN422}
\eeq
The answer has the structure of a sum over saddles with the exponential term being the classical action of the Schwarzian mode and the gauge fields. We sum over winding numbers $m$, $n$ and $r$ to include all the saddles with the same boundary holonomies. The $\U(1)$ charge is quantized as $u \in \mathbb{Z}\cdot u_0$, where $u_0$ should be determined from the microscopic theory. Note also our notation-- $\upalpha$ is a parameter of the theory \eqref{eq:alphagammadef}, whereas $\alpha_\pm$ are chemical potentials which are just our choice of boundary conditions in this ensemble. The one-loop determinant of \eqref{eqn:LN422} is given by
\bea
Z_{\text{1-loop}} &=& \frac{\Phi_r}{\beta}\sqrt{\frac{2\pi(1+\upalpha)^2}{\upalpha}}\underbrace{\frac{(\alpha_++n)}{\sin 2\pi \alpha_+ }\frac{(\alpha_-+m)}{\sin 2\pi \alpha_- }}_{\text{from $\SU(2)$ gauge fields}} \,\times\,\underbrace{4\cos \pi (\alpha_+ + \alpha_-) \cos \pi (\alpha_+ - \alpha_-)}_{\text{from fermion partners of $\U(1)$ mode}}\nonumber\\
&&\times \, \underbrace{\frac{\cos \pi (\alpha_+ + \alpha_-) \cos \pi (\alpha_+ - \alpha_-)}{(1-4(\alpha_+ +\alpha_- + m + n)^2)(1-4(\alpha_+ -\alpha_- - m + n)^2)}}_{\text{from fermions partners of reparam. mode}} \, . \label{eq:Z1looplargeN4}
\ea
 We indicated the origin of the chemical potential-dependent terms in the one-loop determinants. In the computation of the one-loop determinant, we discard all global isometry generators as zero modes from the path integral. The $1/\beta$ factor reflects the difference in the number of bosonic and fermionic zero modes. 

In \cite{Heydeman:2025vcc}, it was shown that the conserved charges of the large $\mathcal{N}=4$ Schwarzian theory satisfy the following algebra\footnote{Note that this is not quite the same as \eqref{eq:commutatorsGG}, \eqref{eq:commutatorsUGQ}, and \eqref{eq:QGVIRASORO}. The relationship between the algebra of conserved charges of the Schwarzian theory and the underlying Virasoro will be reviewed in Section \ref{sec:BPSBHQC}.}
\bea
\label{eq:extralargeRR}
\{ Q_{A\dot{A}}, Q_{B\dot{B}}\} = -\varepsilon_{AB} \varepsilon_{\dot{A}\dot{B}} H \, ,~~~~~\{ \tilde{Q}_{A\dot{A}}, \tilde{Q}_{B\dot{B}}\} = \varepsilon_{AB} \varepsilon_{\dot{A}\dot{B}} \frac{2(1+\upalpha)^2\Phi_r}{\upalpha} \, ,
\ea
and 
\ie\label{QQt}
\{ \tilde{Q}_{A\dot{A}}, Q_{B\dot{B}}\} = \varepsilon_{AB} \varepsilon_{\dot{A}\dot{B}} \i\, U + \varepsilon_{\dot{A}\dot{B}} J_{AB}-\varepsilon_{AB} J_{\dot{A} \dot{B}} \,  .
\fe
 Here $\epsilon_{12}=1$ is the antisymmetric tensor. $Q_{A\dot B}$ are the four fermionic generators associated to $\eta^{A\dot{A}}$, while $\tilde{Q}_{A\dot B}$ are associated to the four free fermions $\psi^{A\dot{A}}$. The $\SU(2)_\pm$ generators are written as $J_A{}^B = (\sigma^I)_{A}{}^B J_+^I$ and $J_{\dot{A}}{}^{\dot{B}} = (\sigma^I)_{\dot{A}}{}^{\dot{B}} J_-^I$, where $\sigma^I$ are the Pauli matrices. Both fermionic generators transform in the bi-fundamental representation of $\SU(2)_+\times\SU(2)_-$, trivially in $\U(1)$, and they commute with $H$. One can show that the unitary bounds among fermionic generators \eqref{eq:extralargeRR} and \eqref{QQt} produce the BPS energy bound in \eqref{eq:N4LBPSEE}.

\smallskip

We derive the spectrum in the microcanonical ensemble. In order to do this we expand the partition function \eqref{eqn:LN422} as
\bea
Z(\beta,\alpha_+,\alpha_-,\mu) &=& \sum_{u \, \in \, u_0 \cdot \mathbb{Z}} \, \, \sum_{j_+, \,j_-} \chi^{\rm long}_{j_+j_-u}(\alpha_+,\alpha_-,\mu)\, \int \d E\, e^{-\beta E} \rho_{j_+j_-u}(E)\nonumber\\
&&~~~~+ \sum_{u\,\in \,u_0 \cdot \mathbb{Z}} \, \, \sum_{j_+, \, j_-} \chi^{\rm short}_{j_+ j_-u}(\alpha_+,\alpha_-,\mu) e^{-\beta E_{\rm BPS}(j_+,j_-,u)} N_{j_+j_-u} \, .\label{eq:ZN4EXP2}
\ea
The two lines of this expansion encode very different parts of the spectrum, one discrete and one continuous. We first focus on the first line, which involves the long 
non-BPS supermultiplet coming with a continuous spectrum. A single long supermultiplet contains a sum over states of different spins; the contribution of this sum to the partition function is the super-character:
\bea
\chi^{\text{long}}_{j_+j_-u} &=& e^{2\pi \i \mu u} \cdot\Big(\chi_{j_+-\frac{3}{2}}\chi_{j_- -\frac{1}{2}} + 2 \chi_{j_+-1}\chi_{j_-}+2\chi_{j_+-1}\chi_{j_--1}+\chi_{j_+-\frac{1}{2}}\chi_{j_-+\frac{1}{2}} \nonumber\\
&&\hspace{-0.3cm}+4 \chi_{j_+-\frac{1}{2}}\chi_{j_--\frac{1}{2}}+\chi_{j_+-\frac{1}{2}}\chi_{j_--\frac{3}{2}}+2\chi_{j_+}\chi_{j_-}+2\chi_{j_+}\chi_{j_--1}+\chi_{j_++\frac{1}{2}}\chi_{j_--\frac{1}{2}}\Big) \, ,\label{eq:XLN4SM}
\ea
where $\chi_{j_\pm}(\alpha_\pm)$ denotes the $\SU(2)_\pm$ character of spin $j_\pm$. The individual states are labeled by their charges under $\SU(2)_+ \times \SU(2)_- \times \U(1)$, while the long supermultiplet structure is generated by the action of $Q_{2\dot 1}$, $Q_{1\dot 1}$ and $\tilde Q_{2\dot 1}$, $\tilde Q_{1\dot 1}$ on a state annihilated by other supercharges. Some multiplets are accidentally shorter when one of the spins or both take the value $1/2$ or $0$ but they still do not preserve any supercharge, see \cite{Heydeman:2025vcc}. A lengthy calculation of the inverse Laplace transform \eqref{eq:ZN4EXP2} using \eqref{eqn:LN422} and \eqref{eq:Z1looplargeN4} reveals these long non-BPS supermultiplets have a positive density of states:
\bea
\rho_{j_+j_-u}(E) &=&  \frac{e^{S_0}\upalpha^{3/2}\,j_+ j_-}{32\Phi_r\sqrt{2\pi}(1+\upalpha)^3}  \left(E-\frac{\upalpha \left((j_+- j_-)^2+u^2\right)}{ 2\Phi_r(1+\upalpha)^2}\right)^{-1}\left(E-\frac{\upalpha \left((j_++ j_-)^2+u^2\right)}{ 2\Phi_r(1+\upalpha)^2}\right)^{-1}\nonumber\\
&&\times \sinh \left(2\pi \sqrt{2\Phi_r \left(E-\frac{\upalpha j_+^2 + j_-^2+\upalpha/(1+\upalpha) u^2}{2\Phi_r(1+\upalpha)}\right)}\right),
\label{eq:NL4DOS}
\ea
which is non-vanishing only for energies above a gap
$$
E\geq E_0(j_+,j_-)= \frac{\upalpha j_+^2 + j_-^2}{2\Phi_r(1+\upalpha)}+ \frac{\upalpha }{2\Phi_r(1+\upalpha)^2}u^2 \, .
$$
One can check that the typical square-root spectral edge becomes an inverse square-root edge for long multiplets satisfying $\upalpha j_+=j_-$. As we see below, this happens when the BPS bound is about to saturate at the edge, and is consistent with the behavior of chaotic supersymmetric quantum mechanics \cite{Stanford:2019vob,Turiaci:2023jfa,Johnson:2024tgg}\footnote{Chaotic theories have a spectrum that shares its statistical properties with that of a random matrix theory without being one. A large class of theories of pure 2d gravity are known to be dual to actual random matrix models. This has not been done for theories with large $\mathcal{N}=4$ supersymmetry, but we are aware of work in progress \cite{WOP_JK} along this direction.}.

In contrast to the long multiplets in the first line of \eqref{eq:ZN4EXP2}, the second line involves the short BPS  multiplets. Their contribution to the partition function is the supercharacter:
\bea
\chi^{\text{short}}_{j_+,j_-,u} &=& e^{2\pi \i \mu u} \cdot\Big(\chi_{j_+} \chi_{j_-+\frac{1}{2}} +2\chi_{j_+-\frac{1}{2}}\chi_{j_-} + \chi_{j_++\frac{1}{2}}\chi_{j_-}\nonumber\\
&&~~+2\chi_{j_+}\chi_{j_--\frac{1}{2}}+\chi_{j_+-1}\chi_{j_--\frac{1}{2}}+\chi_{j_+-\frac{1}{2}}\chi_{j_--1}\Big).\label{eq:XLN4SMA}
\ea
These multiplets are shorter since they are further annihilated by $Q_{2 \dot{1}}$, making them partially BPS. There are even cases where the partial BPS multiplets become even shorter and preserve all four supercharges $Q_{A\dot{A}}$, but still none of the $\tilde Q_{A\dot A}$. This happens when one or both of the $\SU(2)$ spins are zero. For example, the $(j_+,j_-,u)=(j>0,0,u)$ short multiplets are composed of three types of states with $\SU(2)$ spins $(j,1/2)$, $(j-1/2,0)$, and $(j+1/2,0)$, while the $(j_+,j_-,u)=(0,0,u)$ short multiplets are composed of two types of states with spin $(0,1/2)$ and $(1/2,0)$. When $j_+$ or $j_-$ equals $1/2$ the short multiplets can also be accidentally shorter, but they do not preserve more supersymmetry.

It is fairly standard that supersymmetric systems have BPS states such as those counted in \eqref{eq:XLN4SMA}. But a new feature of \eqref{eq:ZN4EXP2} is that the BPS states have a nontrivial energy depending on their charges; they are not all ground states. This energy is given by
\beq
E_{\rm BPS}(j_-,j_+) = \frac{\upalpha}{2\Phi_r(1+\upalpha)^2}\left(\left( j_+ + j_- + \frac{1}{2}\right)^2+ u^2\right).\label{eq:N4LBPSEE}
\eeq 
One can further check that the BPS energy can be reproduced from the unitary bound of the linearized algebra, but the presence of $\Phi_r$ means this is a 1-loop effect in gravity.
The degeneracy of the BPS states appearing in \eqref{eq:ZN4EXP2} also depends on the charges and is given by: 
\beq\label{eqn:NBPS}
N_{j_+j_-u} =\begin{cases}
    e^{S_0} \frac{\sqrt{\pi\upalpha/2} }{32(1+\upalpha)} \sin \left( \pi \frac{2\upalpha j_+ - 2 j_- +\upalpha}{1+\upalpha}\right),~~~~\text{if}   ~~~(j_+,j_-)\in R_{\rm BPS} \, ,\\
    \hspace{0.5cm} 0 \hspace{4.7cm}~~\text{if}   ~~~(j_+,j_-)\notin R_{\rm BPS}\, ,
    \end{cases}
\eeq
where $R_{\rm BPS}$ is a range of charges $(j_+, j_-)$ where BPS multiplets appear in the theory:
\beq
R_{\rm BPS} = \left\{ (j_+,j_-) \left| \, 0< \frac{2\upalpha j_+ - 2 j_- +\upalpha}{1+\upalpha}<1\right.\right\}.
\label{eq:BPSrangeD21}
\eeq
This means that depending on the value of $\upalpha$ and the choice of charges $j_+$, $j_-$ (but not $u$), we may have a large number of BPS states, or none at all! This is illustrated in Figure~\ref{fig:BPSstates}, which compares the spectrum of the Schwarzian theory with the fluctuations of empty AdS$_3$.\footnote{The normalization in \eqref{eqn:NBPS} is chosen to match that of \eqref{eq:Z1looplargeN4}. When applied to higher-dimensional black holes the normalization can be absorbed as a temperature and $j$-independent one-loop correction to $S_0$.}

In the Schwarzian theory, $\upalpha$ is a continuous parameter, while for the string background in section \ref{sec:STI}, varying $\upalpha$ is replaced by discretely changing the number of units of 5-brane flux. We would like to apply the results of the Schwarzian calculation \eqref{eq:NL4DOS}, \eqref{eqn:NBPS} directly to this background, which is undertaken next.

\subsection{Application to BPS black holes in $AdS_3 \times S^3 \times S^3 \times S^1$}\label{sec:BPSBHQC}

In this section, we use the results in Section \ref{sec:LargeN4Review} to evaluate the effect of quantum gravity corrections on the spectrum of near-BPS black holes present in the type II $AdS_3 \times S^3 \times S^3 \times S^1$ background of \ref{sec:STI}. We will find corrections to the near-BPS spectrum of black hole microstates in a similar spirit as the analysis done in \cite{Heydeman:2020hhw} or \cite{Boruch:2022tno}. We also make a prediction for the spectrum of BPS black hole states which is richer than other cases studied so far. Through holography, this also makes a predictions for the spectrum of the 2d CFT dual to the $AdS_3 \times S^3 \times S^3 \times S^1$ background, in the Ramond sector. As we reviewed in the introduction, this CFT is not really known for generic values of the parameters, so this is an interesting situation in which we are using quantum gravity to make a discrete, BPS, microscopic prediction rather than deriving the black hole entropy from the microscopic system.

We shall begin by stating very explicitly the regime in which we will work. First of all, we are interested in the limit where supergravity is an accurate description, so the geometry must be much larger than the size of the fundamental strings. The radii of the spheres is related to the levels, where we take
\beq
k^+,k^- \gg1,~~~~~\text{with fixed }~~k^-/k^+,
\eeq
This also implies that the central charge \eqref{eq:centralcharge} is also very large. Even if gravity is weakly coupled and all curvatures are small in this regime, a key property of near-extremal black holes is that quantum effects can still become large at low enough temperatures (this is most easily seen in the example of AdS$_3$\cite{Ghosh:2019rcj}).

For simplicity, we work in a sector of vanishing $\U(1)$ charges as well as $J_R^+ = J_R^- =0$. Moreover, we will work in an ensemble of fixed, large and positive angular momentum $J$ along $AdS_3$, that is, $\bar{\tau}= \i \beta_R/2\pi$ in the language of \eqref{eq:leftrighttemperatures} with $$\beta_R \sim 2\pi \sqrt{c/24 J} \ll 1.$$ In this limit, the angular velocity is very close to $\Omega \sim -1$. We also work in a low temperature regime with $\tau \sim \i \beta / \pi$, or equivalently $\beta \sim \beta_L/2 $, large. It will be useful to consider $\beta /c$ fixed. This is precisely the regime studied in \cite{Ghosh:2019rcj,Maxfield:2020ale}, with the inclusion of Kac-Moody generators\footnote{While the analysis here is done at the supergravity level, the modular bootstrap analysis of \cite{Ghosh:2019rcj} was refined and made more rigorous recently in \cite{Pal:2023cgk,Dey:2024nje}.}. We are therefore interested in states with large $J$, or large $\bar{h}$ and small $h-c/24 = (H-J)/2$ of order $1/c$.

In the bulk, this regime of parameters leads to a black hole with a long nearly-$AdS_2 \times S^1 \times S^3 \times S^3 \times S^1$ throat with an emergent large $\mathcal{N}=4$ superconformal symmetry. This is straightforward from the analysis of \cite{Ghosh:2019rcj} since the 10d geometry \eqref{eq:BHMET10d} includes a factor with the BTZ metric. AdS/CFT therefore predicts that the boundary theory should also display the same pattern of symmetry breaking, described by the large $\mathcal{N}=4$ Schwarzian theory we reviewed in section \ref{sec:LargeN4Review}. 

The relevant Schwarzian theory has a few parameters $S_0$, $\Phi_r$ and $\upalpha$. The first objective is to identify their values. This can be done at the level of the geometry or, more simply, by comparing the classical free energies. The on-shell action of the 10d black hole \eqref{eq:fullonshellaction} can easily be evaluated in the near-BPS regime, resulting in
\beq
\log Z \approx 2\pi \sqrt{\frac{c J}{6}} + \frac{\pi^2 c}{12 \beta} - \frac{2 \pi^2 k^+}{\beta} \alpha_+^2 - \frac{2\pi^2 k^-}{\beta} \alpha_-^2.
\eeq
This expression can be used to approximate the partition function in the regime of large central charge, and low temperature, but $\beta \lesssim c$. Quantum corrections in Section \ref{sec:LargeN4Review} will allow us to go to lower temperatures and reach the BPS regime. We can use this to extract the relevant parameters
\bea
S_0 &=& 2 \pi \sqrt{\frac{c J}{6}}= 2\pi \sqrt{\frac{Q_1 Q_5^+ Q_5^-}{Q_5^+ + Q_5^-} J},\\
\Phi_r &=& \frac{c}{24} =\frac{Q_1 Q_5^+ Q_5^-}{4(Q_5^+ + Q_5^-)} ,\label{eq:PhirADS3}\\
\gamma &=& \frac{ Q_5^-}{Q_5^+ + Q_5^-}  ,~~~~\upalpha = \frac{Q_5^-}{Q_5^+}.\label{eq:ALPHAADS3} 
\ea
With these identifications we can convert the Schwarzian results into predictions for the spectrum of near-BPS black holes in $AdS_3 \times S^3 \times S^3 \times S^1$. This follows an EFT kind of reasoning; the effective action \eqref{eq:actionbosonicint} is the simplest local expression that is consistent with the pattern of superconformal symmetry breaking exhibited by the UV $AdS_3 \times S^3 \times S^3 \times S^1$ description of the black hole and its near horizon metric fluctuations. But, we still must use the UV description to compute the IR parameters in the EFT.

Before we continue, let us briefly point out how this generalizes if we turn on the right-moving $\SU(2)$ charges. At the classical level, the temperature-dependent part of the action is unchanged. Therefore, the gap scale and the parameter $\upalpha$ are unchanged. The only modification is in the extremal entropy
\beq
S_0 \to 2\pi \sqrt{\frac{ Q_1Q_5^+ Q_5^-}{Q_5^+ + Q_5^-} \Big( J- \frac{\bar{j}_+^2}{Q_1Q_5^+} - \frac{\bar{j}_-^2}{Q_1Q_5^-}\Big)},
\eeq
where $\bar{j}_\pm$ corresponds to the angular momentum along $\SU(2)_R^+ \times \SU(2)_R^-$ directions. The generalization to non-vanishing $\U(1)$ charge is similar. 

To organize the results appropriately, it is useful to recall the supermultiplet structure of 2d CFT with large $\mathcal{N}=4 $ supersymmetry. We focus on the case $A_\gamma$ that is relevant to the background of $AdS_3 \times S^3 \times S^3 \times S^1$. We refer the reader to \cite{Petersen:1989pp,Petersen:1989zz} for details on the representation theory of this algebra. We shall focus on the left-moving sector, since the right-moving analysis is completely analogous. 

\begin{itemize}
\item The non-BPS supermultiplets in the Ramond sector are labeled by the two $\SU(2)_L^+$ and $\SU(2)_L^-$ spins $(j_+,j_-)$. For a given $(j_+,j_-)$, the supermultiplet consists of a set of $\SU(2)^\pm$ Kac-Moody primaries with the same spin spectrum as in Schwarzian theory, shown in equation \eqref{eq:XLN4SM}. The only difference in this regard is that in the 2d CFT case, the spins are bounded by the $\SU(2)$ levels. Since we will work in the regime $k^+,k^-\gg1$ the upper bound will not be important.  All of these states have the same  $\U(1)$ charge $u$. Unitarity implies that these states satisfy the inequality
\beq
h -\frac{c}{24} \geq  \frac{(j_+ +j_- )^2}{k^++k^-}  + \frac{u^2}{k^++k^-} ,
\eeq
where $h$ is the lowest eigenvalue of $L_0$ in the multiplet and is related to the Hamiltonian by $h-c/24 = (H-J)/2$. 

\item The BPS multiplets also have the same spectrum of $\SU(2)_L^+\times \SU(2)_L^-$ spins as the Schwarzian, and we follow the convention in equation \eqref{eq:XLN4SMA}. Their left-moving scaling dimension is fixed to
\beq
h_{\rm BPS}-\frac{c}{24}  = \frac{(j_+ +j_- + 1/2)^2}{k^++k^-}  + \frac{ u^2}{k^++k^-}.  
\eeq
This expression is also equivalent to the Schwarzian counterpart when the spins are appropriately labeled. Notice the non-linear relationship between the BPS scaling dimension and the angular momenta. This will play an important role later when we discuss the index in Section \ref{sec:NEBPSBH}.
\end{itemize}

It is instructive to identify the subset of Virasoro generators that correspond to their Schwarzian counterpart. First of all, since we are interested in black holes, we consider the CFT in the RR sector, so fermionic generators come with integer grading. The relevant subset of the $A_\gamma$ generators and therefore the zero-modes in the Ramond sector 
\beq 
Q^{A\dot{A}}_{\rm Sch} = \sqrt{2}G_0^{A\dot{A}},~~~~J^{AB}_{\rm Sch} = T^{AB}_0,~~~~J^{\dot{A}\dot{B}}_{\rm Sch} = T^{\dot{A}\dot{B}}_0,~~~~H_{\rm Sch} = 2L_0 - \frac{c}{12},
\eeq
while $\tilde{Q}_{\rm Sch}^{A\dot{B}}=Q_0^{A\dot{B}}/\sqrt{2}$ and $U_{\rm Sch} = U_0$. It is easy to check that, with these identifications, the algebra satisfied by the Ramond sector zero modes is equal to the algebra of conserved charges of the Schwarzian theory. For example, in terms of the Schwarzian generators (zero-modes in the Ramond sector) the anticommutator in equation \eqref{eq:QGVIRASORO} reduces to \eqref{QQt}, with the normalization chosen above removing the factors of 2 in \eqref{eq:QGVIRASORO}. We can also check that the anticommutator \eqref{eq:GGVIRASORO} reduces to the left equation in \eqref{eq:extralargeRR} since, when taking the zero-modes, the second line in \eqref{eq:GGVIRASORO} vanishes. Reproducing the right equation in \eqref{eq:extralargeRR} requires a more careful calculation of the coefficient on the RHS since 
\beq
\{ \tilde{Q}^{A\dot{A}}_{\text{Sch}}, \tilde{Q}^{B\dot{B}}_{\text{Sch}}\}= \varepsilon_{AB}\varepsilon_{\dot{A}\dot{B}} \frac{k^++k^-}{2}=\varepsilon_{AB}\varepsilon_{\dot{A}\dot{B}}\frac{2(1+\upalpha)^2 \Phi_r}{\upalpha}, 
\eeq
using the identifications in \eqref{eq:PhirADS3} and \eqref{eq:ALPHAADS3} between the Schwarzian parameters and the $AdS_3 \times S^3 \times S^3 \times S^1$ ones. Finally, \eqref{eq:TGVIRASORO} reduces to the commutation relation between an $\SU(2)$ generator and a bifundamental field since the term involving $Q_m^{A\dot{B}}$ vanishes when the LHS is evaluated for the zero-modes. The rest of the commutation relations are trivial.

In the low-temperature regime, we study the spectrum of $H_{\rm Sch} \approx H-J$, the energy of the state above extremality. We emphasize that although the superconformal algebra $D(2,1|\upalpha)$ appears in the NS sector, it is the algebra of zero modes in the Ramond sector that corresponds to the algebra of conserved charges of the Schwarzian theory. 

We can now assemble the pieces and formulate our prediction for the near-BPS spectrum of the 2d CFT dual to $AdS_3 \times S^3 \times S^3 \times S^1$ in the Ramond sector. We predict the presence of an approximate continuum of states near the BPS bound with a density of states (for $u=0$)
\beq
\rho_{j_+j_-}(h) \approx  e^{2\pi \sqrt{ c J/6}}\, \frac{ j_+ j_-}{2 \sqrt{2} \pi^3 (k^++k^-)^{3/2}}\,\frac{ \sinh \Big(2\pi \sqrt{\frac{k^+k^-}{k^++k^-}\Big(h-\frac{c}{24}- \frac{j_+^2}{k^+} -\frac{j_-^2}{k^-}\Big)}\Big)}{\Big(h-\frac{c}{24}-\frac{(j_+- j_-)^2}{ k^++k^-}\Big)\Big(h-\frac{c}{24}-\frac{ (j_++ j_-)^2}{ k^++k^-}\Big)},\label{eq:NL4DOS3d}
\eeq
with support for energies such that the argument of the square root is positive $h \geq h_0(j_+,j_-)$ with $h_{0}(j_+,j_-)-\frac{c}{24}= \frac{j_+^2}{k^+} + \frac{j_-^2}{k^-}$. (This density of states is defined with respect to a flat measure on $h$.) We can combine this expression with the BPS bound and obtain the gap 
\beq\label{eq:gapRLN4333}
h_{\rm gap}(j_+,j_-) = \frac{k^+k^-}{k^++k^-} \left( \frac{j_+}{k^+} -\frac{j_-}{k^-}\right)^2
\eeq
Therefore the spectrum is very similar to its small $\mathcal{N}=4$ counterpart\cite{Heydeman:2020hhw}, there is a power-law-suppressed gap above the BPS bound, followed by a continuum with a square-root edge that at large enough energies matches the classical Bekenstein-Hawking entropy (the exponential regime of the hyperbolic sine above). This is true except for multiplets satisfying
\beq
j_+ = \frac{k^+}{k^-} j_- \, ,~~~{\rm or}~~~j_+ = \frac{Q_5^+}{Q_5^-} j_- \, , \qquad \textrm{(gapless multiplets)}
\eeq
since for those spins the gap vanishes and the continuum reaches the BPS bound. The appearance of gapless multiplets is similar to the $\mathcal{N}=1$ Schwarzian, or the $\mathcal{N}=2$ Schwarzian with an anomaly~\cite{Stanford:2017thb,Heydeman:2024fgk}. In the present case, the gapless multiplets can only appear when this relation is consistent with angular momentum quantization. This restriction becomes trivial if $k^+=k^-$, or equivalently if the two 3-spheres are equal.

In addition to the above continuous spectrum, the evaluation of the partition function incorporating quantum effects also predicts an exponential number of BPS states and their specific spectrum. In particular, the spectrum of $\SU(2)$ spins is nontrivial. The number of BPS states for the multiplet $(j_+,\, j_-,u)$ is given by equation \eqref{eqn:NBPS} with the appropriate choices of Schwarzian parameters, as indicated in the introduction. The range of angular momenta for which there exist BPS black holes is
\beq
\label{eq:RBPSdefn}
R_{\rm BPS} = \left\{ (j_+,j_-) \left| \, 0< \frac{2 Q_5^- j_+ - 2 Q_5^+ j_- +Q_5^-}{Q_5^+ + Q_5^-}<1\right.\right\}.
\eeq

The fact that there is a special window (which depends on the charges of the states as well as the ``background'' $Q^\pm_5$ parameters) where BPS states occur has some interesting consequences. To see this, as explained in more detail in \cite{Heydeman:2025vcc} for the Schwarzian theory, we can also define the total spin,
\begin{align}
    j_+ +j_-=j \, , \qquad \textrm{(total spin).}
\end{align}
When $Q_5^-/Q_5^+ \to 0$, all BPS states with total spin $j$ accumulate at $(j_+=j,\, j_-=0)$. Since $k^- \sim Q_5^- \sim R_-^2$ in the limit of small $Q_5^-/Q_5^+$, the size of $S_-^3$ is much smaller than the size of the other sphere. In this extreme asymmetric limit, the BPS states are only spherically symmetric on $S_-^3$ and all the angular momentum is on $S_+^3$, the larger sphere. As we increase the 5-brane flux wrapping $S_-^3$, or equivalently, as we increase the size of $S_-^3$, some of the total spin of the BPS state is transferred from $S_+^3$ to $S_-^3$. If we keep $R_+$ fixed for simplicity, this occurs in discrete steps as we increase $R_-$ when we cross
\beq
R_- = \sqrt{\frac{n}{2j_++2j_-+1-n}}\, R_+,~~~n\in\mathbb{Z}, \eeq
and BPS states with spin $( j-\frac{n}{2}, \frac{n}{2})$ disappear and new states with $( j-\frac{n+1}{2}, \frac{n+1}{2})$ appear. What we described in this paragraph is the quantum-corrected version of the classical relation $j_- \approx k_-/k_+ j_+$ derived in \eqref{eq:CLASSJJBPS}. This jumping behavior leads to the discrete points found in Figure \ref{fig:BPSstates}b which are approximated by the smooth classical curves. This jumping behavior is somewhat similar to wall-crossing, although the modulus, the ratio of 5-brane charges, is now discrete. 

We summarize the results of this section with the following points:
\begin{itemize}
    \item In the introduction, we emphasized our result for the spectrum of BPS states in the Ramond sector is considerably richer than the BPS spectrum in the NS sector. As was derived in \cite{Eberhardt:2017fsi}, all BPS states in the NS sector have $j_+ =j_-$ regardless of the relative size between the two three-spheres. This was crucial in order to resolve some puzzles regarding the possible 2d CFT dual. In the Ramond sector as we have just seen, the spin distribution depends on the relative size of the spheres and jumps discontinuously at special values of $Q_5^+$ and $Q_5^-$. 
    \item The spectrum of BPS states is considerably different from its counterpart $AdS_3 \times S^3 \times K3$ or $T^4$. In that case, the BPS bound is simply $h-c/24=0$, and the BPS energies are independent of charges. Moreover, all BPS states would have $j=0$ regardless of the $\SU(2)$ level. There is no adjustable parameter and no jumping of BPS states.
    \item While this result was obtained using the Schwarzian theory (which is only an effective description), the spectrum of BPS states is expected to be more robust beyond the weak coupling regime. So we are really using the gravitational path integral to make a prediction for the microscopic theory. Therefore, it would be interesting to, for instance, understand the origin of these BPS jumps from the point of view of the proposed precise formulation of the dual 2d CFT, either as a symmetric product orbifold or as a sigma model on a certain moduli space of instantons~\cite{Witten:2024yod}.
    \item There is actually an independent way to check some of these results. This is because some aspects of perturbative supergravity on $AdS_3 \times S^3 \times S^3 \times S^1$ are simple enough that they can be directly derived without using the Schwarzian description. This alternative analysis is done in Appendix \ref{app:N4Virasoro}. We derive the one-loop determinant of the large $\mathcal{N}=4$ supergravity mode around the black hole solution above, and show it reproduces the Schwarzian answer; see the discussion around \eqref{eq:app:1loopAdS3}. 
\end{itemize}

Finally, we can repeat the same analysis for the string theory background $AdS_3 \times (S^3 \times S^3 \times S^1)/\mathbb{Z}_2$ analyzed in \cite{Giveon:2003ku,Eberhardt:2018sce}. The orbifold exchanges the two $S^3$ and therefore can only work if $k^- = k^+$. The analysis is very similar to the one performed already and requires the $\mathcal{N}=3$ version of the Schwarzian theory introduced in \cite{Heydeman:2025vcc}.

To explore the BPS states further and construct a protected observable, we can compute an index rather than the partition function. In the next section, we show how to compute the index using the Schwarzian theory, and then independently by a complex saddle-point analysis of the full 10d path integral.

\section{The (modified) elliptic genus from the gravitational path integral}\label{sec:NEBPSBH}

In this section, we consider the gravitational path integral that evaluates the (modified) elliptic genus of $AdS_3 \times S^3 \times S^3 \times S^1$ in the Ramond sector. This is interesting in light of recent developments that clarified the evaluation of supersymmetric quantities using the gravitational path integral \cite{Cabo-Bizet:2018ehj,Iliesiu:2021are}. They involve a geometry that is BPS but non-extremal, although in all examples so far the on-shell action is temperature independent. Curiously, the geometry that evaluates the elliptic genus of $AdS_3 \times S^3 \times S^3 \times S^1$ is BPS, non-extremal, but the on-shell action depends on the temperature. This is expected from the 2d CFT perspective, as the BPS states have a non-vanishing energy \eqref{eq:N4LBPSEE},\footnote{Moreover the energy is non-linear with $R$-charges and therefore the temperature dependence cannot be easily removed via a shift of chemical potentials as in \cite{Cabo-Bizet:2018ehj}.} but interesting to recover from gravity. We can therefore study this protected quantity from both the point of view of the Schwarzian theory, but also this non-extremal, supersymmetric solution.

\subsection{The index of the large $\mathcal{N}=4$ Schwarzian theory}

We begin by describing how to compute the index of the large $\mathcal{N}=4$ Schwarzian theory, following section \ref{sec:LargeN4Review}. More details of this index computation as well as its properties were given in \cite{Heydeman:2025vcc}, and for now we will briefly review this result and take it as a warmup for the calculation for black holes in the 10d $AdS_3 \times S^3 \times S^3 \times S^1$ that we carry out in the rest of the section. 

By choosing a specific value for the chemical potential, and following the notation in \cite{Heydeman:2025vcc}, we can impose supersymmetric boundary conditions with $\alpha_-=\alpha_++1/2$ and define the corresponding Witten index (as a function of $\sigma \equiv \alpha_+$) of the linear large $\mathcal{N}=4$ theory: 
\bea
\text{Index}(\sigma) &\equiv& \frac{1}{4\pi}\frac{\d Z}{\d \alpha_+} (\sigma,\sigma+1/2,\mu)\nonumber
\\
&=& \text{Tr}\left[e^{-\beta H}(-1)^{\sf F}\, J_+^3\, e^{4\pi \i \sigma(J_-^3+J_+^3)} \, e^{2\pi \i \mu U} \right].
\ea
where the fermion parity operator is $(-1)^{\sf F}= e^{2\pi \i J_-^3}$. At $\alpha_-=\alpha_++1/2$, the characters of the long non-BPS multiplets have a second-order zero due to the fermion zero-modes associated to $Q_{2\dot 1}$, $Q_{1\dot 2}$ and $\tilde{Q}_{2\dot 1}$, $\tilde{Q}_{1\dot 2}$, while the short BPS multiplets have a first-order zero from $\tilde{Q}_{2\dot 1}$ and $\tilde{Q}_{1\dot 2}$. The role of the derivative or the insertion of $J_+^3$ is to soak up the fermion zero-modes associated to $\tilde{Q}_{2\dot 1}$ and $\tilde{Q}_{1\dot 2}$, therefore defining an index that receives contributions only from supersymmetry-protected BPS multiplets. This approach which parallels \cite{Gukov:2004fh} is entirely analogous to the helicity supertrace defined to count BPS black holes in flat space \cite{Kiritsis:1997hj}.

It is useful to evaluate the index directly from the path integral approach starting with \eqref{eqn:LN422} and \eqref{eq:Z1looplargeN4} and imposing the supersymmetry relation between chemical potentials. Out of the family of saddles labeled by $m$ and $n$, only those that satisfy $m=n$ and $m=n-1$ are supersymmetric in the bulk as well as in the boundary, and contribute to the path integral. This immediately leads to
\bea
\text{Index}(\sigma)&=& \pi\sqrt{\frac{2\pi}{\beta}\frac{\Phi_r(1+\upalpha)^2}{\upalpha}} \frac{\epsilon}{64\sqrt{\beta/\Phi_r}}\nonumber\\
&&\hspace{-1.1cm}\sum_{n\,\in\, \mathbb{Z}} \sum_{r\, \in\, \mathbb{Z} \cdot u_0^{-1}}  \sum_{\epsilon\, =\, \pm}e^{S_0+\frac{2\pi^2\Phi_r}{\beta}\left(1-4(1+\upalpha)(\sigma +n)^2-4(1+\upalpha^{-1})(\sigma+n-\frac{\epsilon}{2})^2-((1+\upalpha)^2/\upalpha)(\mu+r)^2\right)}\label{eq:index}
\ea
The on-shell action of the saddles is clearly temperature-dependent despite the unbroken supersymmetry, and this is connected to the non-linearity of the BPS bound. Upon a Poisson resummation, this can be written as a sum over states with their $\SU(2)_\pm$ and $\U(1)$ representations
\bea
\text{Index}(\sigma)&=&~\sum_{j \geq 0}\, \, \sum_{u \, \in \, u_0 \cdot \mathbb{Z}}e^{2\pi \i \mu u}\, \sin 2 \pi \sigma \, \chi_{j}(\sigma) \nonumber\\
&&~e^{S_0}\frac{2\sqrt{\pi \upalpha}}{(1+\upalpha)}\,\sin \Big(\frac{\pi (2j+1)}{1+\upalpha}\Big)\, e^{-\beta \frac{\upalpha }{2\Phi_r(1+\upalpha)^2}((j+1/2)^2+u^2)}.
\ea
We can immediately compare this to the full spectrum reviewed in section \ref{sec:LargeN4Review}. First of all, the contributions come solely from short BPS multiplets with $j_-+j_+=j$ as supported by the form of the Boltzmann factor in the equation above. Second, the rest of the terms in the second line are consistent witht the BPS degeneracy of short multiplets with $j_-+j_+=j$ although the index is not enough to determine how the total spin is distributed. This information is unprotected and captured by $R_{\text{BPS}}$ in \eqref{eq:RBPSdefn}. Finally, one can check that the first line is equal to the helicity supertrace of the short BPS multiplets. Therefore, everything is consistent with the spectrum of the large $\mathcal{N}=4$ theory in section \ref{sec:LargeN4Review}.

The goal of this section is to reproduce most of this behavior of the BPS non-extremal geometry contributing to the index, from the 10d black hole solution.

\subsection{The index of the $AdS_3 \times S^3 \times S^3 \times S^1$ black hole}

In section \ref{sec:10dbh}, we found a classical asymptotically $AdS_3 \times S^3 \times S^3 \times S^1$ black hole geometry that evaluates the boundary partition function, for any choice of temperature and angular velocities. Here we motivate the supersymmetric boundary conditions from a 2d CFT analysis and in the next section perform a more detailed derivation of the Killing spinors to verify the guess. 

The supersymmetric partition function of theories with a large $\mathcal{N}=4$ Virasoro symmetry is given by \cite{Gukov:2004fh}
\beq
Z={\rm Tr} \,\left[ e^{-\beta H - \beta \Omega J} \, (-1)^{{\sf F_L}}\, J_L^{-3}\,e^{\beta \Omega^+_L (J_L^{+3}+J_L^{-3})} \, e^{\beta \Omega^+_R J_R^{+3}}\, e^{\beta \Omega^-_R J_R^{-3}} \right].
\eeq
where we used $(-1)^{{\sf F_L}} = e^{2\pi \i J_L^{-3}}$. This supersymmetric partition function is a special case of \eqref{eq:ZTrgravityvars} with
\beq
\Omega_L^-=\Omega_L^+ + \frac{2\pi \i}{\beta}.\label{eq:swwes}
\eeq
The role of the $J_L^{-3}$ insertion is to soak up fermion zero-modes and can be achieved by taking a derivative with respect to $\Omega_L^-$ prior to imposing \eqref{eq:swwes}. In terms of the parameters defined in \eqref{eq:OmegaAlphasss}, the supersymmetry relation is given by
$\alpha_-=\alpha_++1/2$, just like in the Schwarzian theory. This is equivalent to setting two out of the four fermionic generators of the large $\mathcal{N}=4$ to be periodic in the thermal circle.

Having determined that the boundary conditions are supersymmetric we discuss bulk solutions. We can obtain a family of bulk geometries after shifting 
\beq
\alpha_+ \to \alpha_+ + n_+,~~~~~\alpha_- \to \alpha_- + n_-,~~~~~~n_+,n_-\in \mathbb{Z}.
\eeq
This is entirely analogous to the family of saddles in Section \ref{sec:LargeN4Review}. It is important to perform this shifts before imposing the supersymmetry relation between $\alpha_-$ and $\alpha_+$. The on-shell action of these solutions are
\bea
I_{\rm on-shell} &=& -\frac{\i \pi c}{12 \tau} + \frac{2\pi \i k^+}{\tau} \alpha_+^2 + \frac{2\pi \i k^-}{\tau} \alpha_-^2 + \text{h.c.},\nonumber\\
&\to& -\frac{\i \pi c}{12 \tau} + \frac{2\pi \i k^+}{\tau} (\alpha_++n_+)^2 + \frac{2\pi \i k^-}{\tau} (\alpha_++1/2+n_-)^2 + \text{h.c.}
\ea
In the second line, we applied the shift and specialized $\alpha_-=\alpha_++1/2$. 

Not all saddles that have supersymmetric boundary conditions preserve supersymmetry in the bulk.\footnote{This is similar to what happens for the Reissner-Nordstrom black hole, only the saddle with $\Omega = 2\pi \i /\beta$ is supersymmetric \cite{Iliesiu:2021are}.} Only the saddles with $n_+ = n_-$ or $n_+ = n_- + 1$ have globally defined Killing spinors. In the rest of this section we verify this by explicitly constructing the Killing spinors. We also verify this in Appendix \ref{app:N4Virasoro} by explicitly showing that for these saddles the one-loop determinant does not vanish, which can only happen in the presence of a Killing spinor.

The on-shell action of the solutions that do contribute to the modified elliptic genus with $\alpha_-=\alpha_++1/2$ is
\bea
I_{\rm on-shell}= -\frac{\i \pi c}{12 \tau} + \frac{2\pi \i k^+}{\tau} (\alpha_++n)^2 + \frac{2\pi \i k^-}{\tau} (\alpha_++n\pm 1/2)^2 + \text{h.c.},
\ea
where $n=n_+$. The upper sign corresponds to $n_-=n_+$ and the lower sign to $n_+=n_-+1$. The resulting action, and therefore the partition function, depends not only on the right-moving temperature $\bar{\tau}$ (as expected in analogy to the small $\mathcal{N}=4$ elliptic genus) but also depends on the left-moving temperature $\tau$. In the near-BPS regime of the path integral, we obtain a temperature-dependent index just as we did for the Schwarzian theory in \cite{Heydeman:2025vcc}.

We did not worry about the $\U(1)$ mode so far since we are fixing its charge. This analysis could be generalized to fixed potential, but either way, we do need to include quantum fluctuations of the extra fermions $Q^{A\dot{B}}_n$. These fermions actually render the index defined above zero, if it wasn't for the $J_L^{-3}$ insertion \cite{Gukov:2004ym}.  This is a 1-loop quantum effect around the supersymmetric background, while so far we mostly discussed classical aspects of the non-extremal BPS solution. A more explicit discussion is in Appendix \ref{app:N4Virasoro} where we evaluate the one-loop determinant.

The index above depends on the (left-moving) temperature for the same reason as the Schwarzian index depends on temperature. The crucial point is that the BPS bound involves the $R$-charges in a non-linear fashion. Therefore, there is no shift of the angular velocities that could remove the temperature dependence. Such a shift is possible when the BPS bounds are linear with the $R$-charges, as illustrated in \cite{Cabo-Bizet:2018ehj}. In our case, removing the $\tau$ dependence of the index would require turning on a source for a double-trace operator quadratic in the angular momentum along the spheres. Although this might be possible, it seems to us highly impractical, especially from the point of view of the gravitational path integral.

\subsection{Killing spinors of the non-extremal BPS black hole}
In the previous section, we have discussed a complex black hole solution in which the fermion parity operator is implemented by an imaginary angular velocity via the relation $(-1)^{{\sf F_L}} = e^{2\pi \i J_L^{-3}}$. Abstractly, this solution implements supersymmetric boundary conditions for fermions but remains at finite temperature, similar to the complex black hole saddle for AdS$_5$ $\times$ $S^5$ and 4d Reissner-Nordstrom. The goal of this subsection is to explicitly construct the Killing spinors for this large $\mathcal{N}=4$ solution, thereby demonstrating the finite temperature 10-dimensional geometry preserves some supersymmetry. This analysis reveals the algebra of preserved charges and confirms that the black holes transform in the correct short representations of large $\mathcal{N}=4$ supersymmetry.

We begin with the generic form of the extended BTZ solution which has rotations in both the AdS$_3$ and Cartan directions of $S^3_\pm$:
\bea
\label{eq:susyBTZmetric}
\d s^2 &=& - f \d t^2 + \frac{\ell^2 \d r^2}{f} + r^2 \Big[\d \varphi  - \frac{r_- r_+ }{r^2} \d t\Big]^2 \nonumber\\
&&+R_+^2 \Big[\d \theta_+^2 + \sin^2 \theta_+ \Big(\d \phi_+ +  \frac{A_L^{+\,3}+A_R^{+\,3}}{2}\Big)^2 +\cos^2 \theta_+ \Big(\d \psi_+ + \frac{A_L^{+\,3}-A_R^{+\,3}}{2}\Big)^2\Big] \nonumber\\
&& + R_-^2 \Big[\d \theta_-^2 + \sin^2 \theta_- \Big(\d \phi_- +  \frac{A_L^{-\,3}+A_R^{-\,3}}{2}\Big)^2 +\cos^2 \theta_- \Big(\d \psi_- + \frac{A_L^{-\,3}-A_R^{-\,3}}{2}\Big)^2\Big]\nonumber\\
&& + \, L^2 \d \theta^2 \, , \\
f&=&(r^2-r_+^2)(r^2-r_-^2)/r^2 \, , \\
A_L^{\pm 3} &=& 2 \Omega^\pm_L\,   \frac{\d \varphi - \Omega \d t}{1-\Omega},~~~~~A_R^{\pm 3} = 2 \Omega^\pm_R  \, \frac{\d \varphi - \Omega \d t}{1+\Omega} \, .
\ea

To simplify the calculation of the Killing spinors, we will pass to rotating coordinates on the two $S^3$'s, as defined by the 1-forms:
\begin{align}
\label{eq:susyBTZrotcords}
    \d \tilde{\phi}_+ &\equiv \d \phi_+ +  \frac{A_L^{+\,3}+A_R^{+\,3}}{2} \, , \qquad \d \tilde{\psi}_+ \equiv \d \psi_+ + \frac{A_L^{+\,3}-A_R^{+\,3}}{2}\, , \\
   \d \tilde{\phi}_- &\equiv \d \phi_- +  \frac{A_L^{-\,3}+A_R^{-\,3}}{2}\, , \qquad \d \tilde{\psi}_- \equiv \d \psi_- + \frac{A_L^{-\,3}-A_R^{-\,3}}{2} \, .
\end{align}
In terms of the $(\tilde{\phi}_+, \tilde{\psi}_+, \tilde{\phi}_-,\tilde{\psi}_-)$ coordinates, the $S^3\times S^3$ metric is the round one. The form of the 10d IIB RR 3-form in these coordinates is:
\begin{align}
\label{eq:susyBTZflux}
F_3 = 2 r\,  (\d t \wedge \d r \wedge \d \varphi) +  R_+^2 \sin 2\theta_+  (\d \theta_+ \wedge \d \tilde{\phi}_+ \wedge \d \tilde{\psi}_+)  \, \, +  R_-^2 \sin 2\theta_-   (\d \theta_- \wedge \d \tilde{\phi}_- \wedge \d \tilde{\psi}_-) \,   ,
\end{align}
Here and throughout, we have set the coupling $g=1$ appearing in \eqref{eq:typeIIBaction} for simplicity.

The solution we have presented above has the feature that the angular velocities $(\Omega_L^\pm, \Omega_R^\pm)$ do not appear anywhere in the metric. Instead, they will appear as boundary conditions for various fields around the non-contractible cycle (the angular velocity in AdS$_3$ is still present through the relation $\Omega = \frac{r_-}{r_+}$). For generic values of the angular velocities, we always have a solution, but not necessarily one with supersymmetric boundary conditions. As outlined in the previous section \eqref{eq:swwes}, the implementation of supersymmetric boundary conditions is:
\begin{align}
\label{eq:killingspinorboundaryconds}
\Omega_L^- = \Omega_L^+ + \frac{2 \pi \i}{\beta} \, \, , \quad \textrm{(Supersymmetric boundary condition)}\, ,
\end{align}
whereas $\Omega_R^\pm$ remains unconstrained and can take any value while preserving supersymmetry. These global conditions (which generally imply a complex metric) will be important later, after determining the local solution to the Killing spinor equation.

The metric \eqref{eq:susyBTZmetric} together with the RR 3-form \eqref{eq:susyBTZflux} preserves some of the original 32 supersymmetries of Type IIB supergravity if one may find covariant constant Killing spinors $\varepsilon$ which are annihilated by the supersymmetry transformations of the theory. The number of such spinors determine the number of unbroken supersymmetries, and bilinears constructed from these Killing spinors are automatically Killing vectors \cite{Gauntlett:1998kc,Gauntlett:1998fz,Figueroa-OFarrill:1999klq}. Their algebra determines the algebra of unbroken supersymmetries. Because all fermions vanish in the classical solution, the bosonic variations automatically vanish and we need only check the supersymmetry variation of the Type IIB fermions. Type IIB is a chiral theory, so in our conventions all 10-dimensional Majorana-Weyl fermions come in SO(2) doublets of the same chirality, indicated here by the index $i=1,2$. In describing these fermions and their supersymmetry transformation rules, we will follow the conventions of \cite{Ortin:2015hya}. As there, the chiralities of the gravitino, dilatino, and supersymmetry parameter are, respectively,
\begin{align}
\label{eq:chiralitycond}
    \Gamma_{11} \psi^i_\mu = -\psi^i_\mu \, , \qquad \Gamma_{11} \lambda^i = \lambda^i \, , \qquad \Gamma_{11} \varepsilon^i = - \varepsilon^i \, .
\end{align}
Given that our background only contains a nontrivial metric and $F_3$ flux (dual to the magnetic $F_7$ flux), and we have set a constant dilaton and $g=1$ for simplicity, the supersymmetry transformations for the fermions reduce to
\begin{align}
\label{eq:BTZsusyvariations}
    \delta_\varepsilon \psi^i_\mu &= \left(\partial_\mu + \frac14 \omega^{ab}_\mu \Gamma_{ab}\right) \varepsilon^i +\frac{1}{16} \left( \frac{1}{3!} \slashed{F}_3 \Gamma_\mu (\sigma_1)^i_j + \frac{1}{7!} \slashed{F}_7 \Gamma_\mu (\sigma_1)^i_j\right)\varepsilon^j \, \\
    \delta_\varepsilon \lambda^i & = \frac{1}{4} \left( -\frac{1}{3!} \slashed{F}_3  (\sigma_1)^i_j + \frac{1}{7!} \slashed{F}_7  (\sigma_1)^i_j\right)\varepsilon^j\, .
    \label{eq:BTZsusyvariations2}
\end{align}
In local coordinates with Minkowski metric $\eta_{ab}$, we have introduced Gamma matrices which satisfy the anticommutation relations
\begin{equation}
\label{eq:localGammaAlgebra}
    \{ \Gamma_a, \Gamma_b \} = 2 \eta_{ab} \, .
\end{equation}
At this point, we will not need to pick a specific representation of these matrices. Eventually, we will use the conventions of Appendix \ref{app:sugraconventions} to write the explicit solution.

Our choice of rotating coordinates, \eqref{eq:susyBTZrotcords} has reduced the complexity of the metric \eqref{eq:susyBTZmetric}, but the resulting Killing spinor equations are still somewhat complicated and result in spinors which depend on the angles of $S^3_\pm$. We introduce a less obvious frame field associated to \eqref{eq:susyBTZmetric}, and as in \cite{Cabo-Bizet:2018ehj} this will result in a simpler form for the Killing spinors after one determines the spin connection from the frame in the usual way. Our choice of frame is
\begin{align}
\label{eq:SUSYBTZframe}
    e^0 &= \sqrt{f(r)} \, \d t \, , \qquad e^1 =  \frac{\ell}{\sqrt{f(r)}}\, \d r \, , \qquad e^2 =  r(\d \varphi - \frac{r_+ r_-}{r^2}\d t)\, , \\
    e^3 &=  R_+ \left ( \cos(\tilde{\phi}_+ + \tilde{\psi}_+) \d \theta_+ + \frac12 \sin(2 \theta_+) \sin(\tilde{\phi}_+ + \tilde{\psi}_+)(\d \tilde{\psi}_+ - \d \tilde{\phi}_+ )\right )\, , \\
    e^4 &= R_+ \left ( -\sin(\tilde{\phi}_+ + \tilde{\psi}_+) \d \theta_+ + \frac12 \sin(2 \theta_+) \cos(\tilde{\phi}_+ + \tilde{\psi}_+)(\d \tilde{\psi}_+ - \d \tilde{\phi}_+ )\right ) \, , \\
    e^5 &=  R_+ \left ( \sin^2(\theta_+) \d \tilde{\phi}_+ + \cos^2(\theta_+) \d \tilde{\psi}_+\right )\, , \\
    e^6 &=  R_- \left ( \cos(\tilde{\phi}_- + \tilde{\psi}_-) \d \theta_- + \frac12 \sin(2 \theta_-) \sin(\tilde{\phi}_- + \tilde{\psi}_-)(\d \tilde{\psi}_- - \d \tilde{\phi}_- )\right )\, , \\
    e^7 &= R_- \left ( -\sin(\tilde{\phi}_- + \tilde{\psi}_-) \d \theta_- + \frac12 \sin(2 \theta_-) \cos(\tilde{\phi}_- + \tilde{\psi}_-)(\d \tilde{\psi}_- - \d \tilde{\phi}_- )\right ) \, , \\
    e^8 &=  R_- \left ( \sin^2(\theta_-) \d \tilde{\phi}_- + \cos^2(\theta_-) \d \tilde{\psi}_-\right )\, , \\
    e^9 &=  L \,  \d \theta\, .
\end{align}
This frame may be used to determine the components of the spin connection $\omega_\mu^{ab}$ appearing in \eqref{eq:BTZsusyvariations}; there are many nonzero components in this frame but our choices will result in simplifications later. One may further evaluate the components of $F_3$ given in \eqref{eq:susyBTZflux} in this frame, leading to the important matrix:
\begin{align}
\label{eq:F3slashed}
    \slashed{F}_3 \equiv (F_3)^{abc}\Gamma_{abc} = 12 \left ( -\frac{\Gamma_{012}}{\ell} + \frac{\Gamma_{345}}{R_+} + \frac{\Gamma_{678}}{R_-}\right ) \equiv 12 P_\Gamma \, ,
\end{align}
where we have defined $P_\Gamma$, thought of as a projector matrix. Our discussion in what follows is somewhat similar to \cite{Gauntlett:1998kc}, which discusses the vacuum AdS$_3$ solution from the M-theory viewpoint. In our case, we have the complexified BTZ solution as an RR Type IIB background.

By dualizing $F_7$ appearing in the general transformation rules \eqref{eq:BTZsusyvariations2} and using the fact that $\varepsilon$ has negative chirality under $\Gamma_{11}$, the dilatino variation may be written simply in terms of $P_\Gamma$ \eqref{eq:F3slashed}. The condition for unbroken supersymmetry coming from the dilatino is then:
\begin{align}
\label{eq:dilatinosusy}
    \delta_\varepsilon \lambda^i & = - P_\Gamma  \, (\sigma_1)^i_j \varepsilon^j = 0 \, .
\end{align}
To solve this equation for some Killing spinors $\varepsilon^i$, we further introduce
\begin{align}
\label{eq:BigGammaProject}
    \Gamma \equiv \left( \frac{\ell}{R_+} \Gamma_{012345} + \frac{\ell}{R_-} \Gamma_{012678} \right ) \, .
\end{align}
This satisfies
\begin{align}
\label{eq:10dprojector}
    P_\Gamma \Gamma = -P_\Gamma \, , \qquad \Gamma^2 = 1 \, .
\end{align}
Therefore, it follows that any spinor $\varepsilon$ that satisfies the eigenspace equation,
\begin{align}
\label{eq:susyprojection}
    \Gamma \varepsilon^i =  \varepsilon^i \, ,
\end{align}
also satisfies
\begin{align}
    P_\Gamma \varepsilon^i = 0 \, ,
\end{align}
so that \eqref{eq:dilatinosusy} is solved for any spinor which obeys \eqref{eq:susyprojection}. This projection condition breaks half of the supersymmetry, so 16 spinors survive at this step.

To solve the gravitino equations, we can write \eqref{eq:BTZsusyvariations} as
\begin{align}
\label{eq:BTZsusymatrix}
    \delta_\varepsilon \psi^i_\mu &= \partial_\mu \varepsilon^i + \frac14 e^a_\mu \left ( \omega^{bc}_a \Gamma_{bc} \varepsilon^i +  P_\Gamma \Gamma_a (\sigma_1)^i_j \varepsilon^j \right ) \equiv \partial_\mu \varepsilon^i + e^a_\mu  (M_a)^i_j \varepsilon^j= 0 \, .
\end{align}
This defines the matrix $(M_a)^i_j$ whose component expressions can be determined from the various frames in \eqref{eq:SUSYBTZframe}. To begin constructing the solution, we can first require that the Killing spinor is independent of the internal $S^1$ direction, such that $\partial_\theta \varepsilon^i = 0$. This means that the 10th component of \eqref{eq:BTZsusymatrix} is an algebraic equation, and it reduces to our previous condition since $\omega_\theta^{ab} = 0$ and $\{ P_\Gamma, \Gamma_9 \} = 0$.

Along the $S^3_+$ directions labeled by the frame index $a=3,4,5$, the matrix $M_a$ reads:
\begin{align}
    (M_3)^i_j &= -\frac{\Gamma_{45}}{2 R_+}\delta^i_j + \frac14 P_\Gamma \Gamma_3 (\sigma_1)^i_j \, , \\ 
    (M_4)^i_j &= \frac{\Gamma_{35}}{2 R_+}\delta^i_j + \frac14 P_\Gamma \Gamma_4 (\sigma_1)^i_j\, ,\\ 
    (M_5)^i_j &= -\frac{\Gamma_{34}}{2 R_+}\delta^i_j + \frac14 P_\Gamma \Gamma_5 (\sigma_1)^i_j \, .
\end{align}
We want to use the condition $P_\Gamma \varepsilon = 0$, so we commute $P_\Gamma$ to the right, so that when acting on such constrained spinors, we have
\begin{align}
    (M_3)^i_j \varepsilon^j &= -\frac{\Gamma_{45}}{2 R_+}(\delta^i_j - (\sigma_1)^i_j)\varepsilon^j \, ,\\ 
    (M_4)^i_j \varepsilon^j&= \frac{\Gamma_{35}}{2 R_+}(\delta^i_j - (\sigma_1)^i_j)\varepsilon^j\, ,\\ 
    (M_5)^i_j \varepsilon^j&= -\frac{\Gamma_{34}}{2 R_+}(\delta^i_j - (\sigma_1)^i_j)\varepsilon^j\, .
\end{align}
In matrix notation which suppresses the $SO(2)$ vector notation, these components vanish if the spinor $\varepsilon$ further satisfies the projection condition
\begin{align}
    \frac12 (1 - \sigma_1) \varepsilon = 0 \, .
\end{align}
An identical analysis for $S^3_-$ shows that $M_5$, $M_6$, $M_7$ also vanish for a spinor subject to the same two conditions.

In summary, for a Killing spinor which satisfies
\begin{align}
\label{eq:SigmaGammaEigen}
    \sigma_1 \varepsilon = \varepsilon \, , \qquad \Gamma \varepsilon =  \varepsilon \, ,
\end{align}
the Killing spinor equation \eqref{eq:BTZsusymatrix} requires this spinor to be constant on the $S^3 \times S^3 \times S^1$ directions\footnote{Recall that spinor fields transform under local Lorentz transformations that leave the metric invariant; here we have chosen a particularly simple gauge which leaves only spacetime dependence inside AdS$_3$. The more obvious choice of frame will lead to the familiar spinors which depend on the angles of $S^3_\pm$.}. Of the original 32 spinors available in Type IIB, these conditions reduce the number of free components to 8.

Having removed all dependence on the internal sphere directions, it remains to solve the remaining Killing spinor equations for the supersymmetric BTZ, and then use the supersymmetric boundary condition \eqref{eq:killingspinorboundaryconds}. For the components along the AdS$_3$ directions with $a=0,1,2$, the matrix $M_a$ acting on a Killing spinor reads:
\begin{align}
    (M_0)^i_j \varepsilon^j &= \left ( \frac{f'(r)}{4 \ell \sqrt{f(r)}}\Gamma_{01} \delta^i_j-\frac{\Gamma_{12} }{2 \ell r^2}(r_+ r_- \delta^i_j -r^2 (\sigma_1)^i_j) \right) \varepsilon^j \, ,\\ 
    (M_1)^i_j \varepsilon^j&= -\frac{\Gamma_{02}}{2 \ell r^2}\left(r_+ r_- \delta^i_j - r^2 (\sigma_1)^i_j \right)\varepsilon^j\, ,\\ 
    (M_2)^i_j \varepsilon^j&= - \frac{1}{2 \ell r^2} \left ( r \sqrt{f(r)} \Gamma_{12}\delta^i_j + \Gamma_{01}(r_+ r_- \delta^i_j + r^2 (\sigma_1)^i_j) \right )\varepsilon^j\, .
\end{align}
On the eigenspace spanned by \eqref{eq:SigmaGammaEigen}, the Killing spinor equations \eqref{eq:BTZsusymatrix} become:
\begin{align}
     (\partial_t  &+ \frac{(r^2 + r_+ r_-)}{2 \ell r}\Gamma_{01} + \frac{\sqrt{(r^2-r_+^2)(r^2-r_-^2)}}{2\ell r}\Gamma_{12}  ) \varepsilon^i =0 \, , \\
   ( \partial_r   &+ \frac{(r^2 - r_+ r_-)}{2r \sqrt{(r^2-r_+^2)(r^2-r_-^2)}} \Gamma_{02}   ) \varepsilon^i = 0\, , \\
   ( \partial_\varphi &- \frac{(r^2 + r_+ r_-)}{2 \ell r}\Gamma_{01} - \frac{\sqrt{(r^2-r_+^2)(r^2-r_-^2)}}{2\ell r}\Gamma_{12}  ) \varepsilon^i = 0\, .
\end{align}
Our solution strategy now follows that of \cite{Coussaert:1993jp,Giribet:2024nwg}, and in particular we will make direct contact with \cite{Giribet:2024nwg} using the change of variable:
\begin{align}
    J = \frac{2 r_+ r_-}{\ell}\, , \quad \tilde{t} = \ell t \, , \quad \tilde{f}=\frac{1}{\ell}\sqrt{f} \, .
\end{align}
To write an explicit solution, we will pick conventions for the 10-dimensional Clifford algebra as in Appendix \ref{app:sugraconventions}. The 3-dimensional subalgebra can be repackaged as: 
\begin{align}
    \Gamma_0 \Gamma_1 \equiv -\sigma_3 \otimes \mathbf{1}_2 \otimes \mathbf{1}_8 \, , \quad \Gamma_1 \Gamma_2 \equiv \i \sigma_1 \otimes \sigma_1 \otimes \mathbf{1}_8 \, , \quad \Gamma_0 \Gamma_2 \equiv \sigma_2 \otimes \sigma_1 \otimes \mathbf{1}_8 \, ,
\end{align}
where our conventions in Appendix \ref{app:sugraconventions} were adapted to give a four-dimensional version of the choices made in \cite{Giribet:2024nwg} for the $AdS_3$ components of the spinor. After this, our system becomes,
\begin{align}
     (\partial_{\tilde t}  &+ \frac{1}{2\ell}\left [\tilde{f} \Gamma_{12} + \Gamma_{01} (\frac{J}{2r} + \frac{r}{\ell})  \right ] ) \varepsilon^i =0 \, , \label{eq:KStilde1}\\
   ( \partial_r   &- \frac{1}{2 r \tilde{f}}\Gamma_{02}(\frac{J}{2r} - \frac{r}{\ell})   ) \varepsilon^i = 0\, , \label{eq:KStilde2}\\
   ( \partial_\varphi &- \frac{1}{2}\left [\tilde{f} \Gamma_{12} + \Gamma_{01} (\frac{J}{2r} + \frac{r}{\ell})  \right ]  ) \varepsilon^i = 0\, , \label{eq:KStilde3}
\end{align}
which is a small modification of the $\theta=1$ case of Eq. 2.14 in \cite{Giribet:2024nwg}. We can now introduce the left and right moving coordinates (which are slightly different than those in \eqref{eq:lightconecords}),
\begin{align}
    x^\pm = \tilde{t} \pm \ell \varphi \, .
\end{align}
In these coordinates, the Killing spinor equations \eqref{eq:KStilde1}, \eqref{eq:KStilde2}, and \eqref{eq:KStilde3} reduce to a pair of equations, as the spinor becomes independent of $x^+$:
\begin{align}
    \partial_{+}\varepsilon^i = 0 \, .
\end{align}
One could now follow the same techniques in \cite{Giribet:2024nwg}, which shows that the solution along the AdS$_3$ factor is a sum of spinors with fixed positive and negative frequencies in $x^-$. However, the most general solution from this technique will not satisfy our conditions for supersymmetry from higher dimensions, \eqref{eq:SigmaGammaEigen}. Additionally, due to the complex angular velocity (which has been removed locally from the metric but still influences the global properties of the spinor), there are additional supersymmetry conditions \eqref{eq:gamma3467proj} we will encounter after taking care of global considerations. 

To write the general solution subject to the supersymmetry constraints, we will work with eigenstates of definite spin under $\Gamma_{01}$, $\Gamma_{34}$, $\Gamma_{67}$. These matrices commute with $\Gamma_{11}$ and $\Gamma =\left( \frac{\ell}{R_+} \Gamma_{012345} + \frac{\ell}{R_-} \Gamma_{012678} \right )$, so it is consistent to impose these conditions simultaneously. Therefore, similar to \cite{Gauntlett:1998kc}, we will introduce constant spinors $\chi$ which may be decomposed as:
\begin{align}
\label{eq:genconstantspinor}
    \chi = \left ( \chi_{+++} + \chi_{++-} +\chi_{+-+} +\chi_{-++} + \chi_{+--} + \chi_{-+-} + \chi_{--+}+ \chi_{---}\right ) \, , \\ 
    \Gamma_{01} \chi_{\pm \, \cdot \, \cdot } = \pm \chi_{\pm \, \cdot \, \cdot} \,  \, , \qquad \Gamma_{34} \chi_{\cdot \, \pm \, \cdot } = \pm \i \chi_{\cdot \, \pm \, \cdot } \, \, , \qquad \Gamma_{67} \chi_{\cdot \, \cdot \, \pm } = \pm \i \chi_{\cdot \, \cdot \, \pm } \, 
\end{align}
We can now impose our projection conditions \eqref{eq:chiralitycond}, \eqref{eq:SigmaGammaEigen} (and as we see later, \eqref{eq:SigmaGammaOmegaEigen}) on \eqref{eq:genconstantspinor}. The components that survive are:
\begin{align}
\label{eq:constrainedconstantspinor}
    &\chi = \left ( \chi_{++-} + \chi_{+-+} + \chi_{-+-} + \chi_{--+}\right ) \, , \\
    &\Gamma \chi_{++-} = \chi_{++-} \, , \quad \Gamma \chi_{+-+} = \chi_{+-+} \, , \quad \Gamma \chi_{-+-} = \chi_{-+-} \, , \quad \Gamma \chi_{--+} = \chi_{--+} \, .
\end{align}
Note that the spins on the $S^3_+ \times S^3_-$ are anti-correlated as a consequence of \eqref{eq:SigmaGammaOmegaEigen}. Each factor is separately chiral with respect to the 10-dimensional parity matrix, and each is a positive eigenstate of \eqref{eq:BigGammaProject} as we have indicated above. Imposing this projector explicitly leads to more complicated component expressions because it involves $(\ell, R_+, R_-)$; we provide the component expressions for completeness in Appendix \ref{app:sugraconventions}.

We can now write the full solution to \eqref{eq:KStilde1}, \eqref{eq:KStilde2}, and \eqref{eq:KStilde3} subject to the projection conditions. The solution is:
\begin{align}
    \varepsilon^i =& e^{-\frac{\i }{2 \ell} \omega x^-} \left(\sqrt{\frac{r}{\ell} + \frac{J}{2r} - \i \omega} \, \, \chi_{++-} + \sqrt{\frac{r}{\ell} + \frac{J}{2r} + \i \omega}  \, \, \chi_{-+-} \right)\eta^i_1 \nonumber \\ 
&\!\!\! \! + e^{\frac{\i }{2 \ell} \omega x^-}\left(\sqrt{\frac{r}{\ell} + \frac{J}{2r} + \i \omega}  \, \, \chi_{+-+} + \sqrt{\frac{r}{\ell} + \frac{J}{2r} - \i \omega} \, \, \chi_{--+} \right)\eta^i_2
\end{align}
where,
\begin{align}
    \omega = \sqrt{-\frac{J}{\ell}-M} \, , \qquad J = \frac{2 r_+ r_-}{\ell} \, , \qquad M = \frac{r_+^2 + r_-^2}{\ell^2} \, ,
\end{align}
and $\eta_1^i$, $\eta_2^i$ are constants. It is also helpful to recast the solution in terms of our original variables, 
\begin{align}
\label{eq:KSfiniteT}
    \varepsilon^i =& \frac{e^{\frac{(r_+ + r_-)}{2 \ell} (t-\varphi)}}{\sqrt{\ell r}} \left(\sqrt{(r+r_+)(r+r_-)} \, \, \chi_{++-} + \sqrt{(r-r_+)(r-r_-)}  \, \, \chi_{-+-} \right)\eta^i_1 \nonumber \\ 
&\!\!\! \! + \frac{e^{-\frac{(r_+ + r_-)}{2 \ell} (t-\varphi)}}{\sqrt{\ell r}}\left(\sqrt{(r-r_+)(r-r_-)}  \, \, \chi_{+-+} + \sqrt{(r+r_+)(r+r_-)} \, \, \chi_{--+} \right)\eta^i_2 \, .
\end{align}
It remains to verify that this spinor satisfies the correct boundary conditions. In doing so, we will also verify the number of supersymmetries preserved by this solution.

\subsection*{Temperature dependence and boundary conditions}
The local solution of the Killing spinor equations was determined to be \eqref{eq:KSfiniteT}. We still must verify that this spinor is globally well defined and has the correct supersymmetric boundary conditions. The factors in the exponent can be written explicitly in terms of $\beta=2\pi \ell r_+/(r_+^2-r_-^2)$ and $\Omega=r_-/r_+$, demonstrating that we have found a Killing spinor that depends on the temperature-- this will ultimately be possible due to the imaginary rotations used to implement the index in gravity. The relationship is: 
\begin{align}
\label{eq:KSphase}
    \frac{(r_+ + r_-)}{2 \ell} = \frac{\pi}{\beta(1-\Omega)} \, .
\end{align}

Now, a crucial point is that the Killing spinor is globally well defined only if it has antiperiodic boundary conditions around the contractible cycle. The may still be twisted around the contractible cycle due to our coordinate choice \eqref{eq:susyBTZrotcords} which absorbed the $SU(2)_+ \times SU(2)_-$ angular momentum. In summary, we would like to demand:
\begin{align}
    (t,\varphi) &\sim (t-\i \beta, \varphi-\i \beta \Omega) \, , \quad \textrm{(contractible, antiperiodic spinor)} \, .
\end{align}
Going around the contractible cycle, we have from \eqref{eq:KSfiniteT} and \eqref{eq:KSphase},
\begin{align}
    e^{\pm \frac{(r_+ + r_-)}{2 \ell} (t-\varphi)} = e^{\pm \frac{\pi}{\beta(1-\Omega)} (t-\varphi)} \longrightarrow e^{\pm \frac{\pi}{\beta(1-\Omega)} (t-\varphi - \i \beta (1-\Omega))} = - e^{\pm \frac{\pi}{\beta(1-\Omega)} (t-\varphi)} \, .
\end{align}
Therefore, the spinor is globally well defined. Note that the spinor does not depend on the angles $\tilde{\phi}_\pm$ and $\tilde{\psi}_\pm$, and also that the rotating coordinates \eqref{eq:susyBTZrotcords} are invariant around this cycle. 

We next check the periodicity around the non-contractible cycle:
\begin{align}
\label{eq:KSnoncontphase}
    e^{\pm \frac{\pi}{\beta(1-\Omega)} (t-\varphi)} \longrightarrow e^{\pm \frac{\pi}{\beta(1-\Omega)} (t-\varphi -2\pi)} =  e^{\pm \frac{\pi}{\beta(1-\Omega)} (t-\varphi)}e^{\mp \frac{2\pi^2}{\beta(1-\Omega)}} \, .
\end{align}
Naively, this implies the spinor is not single valued on the spatial circle. However, here it is important that we solved for the Killing spinor only in our special frame and coordinate system \eqref{eq:susyBTZrotcords}. The rotating coordinates in particular mean that we performed a gauge transformation to remove those terms from the metric, namely the components $(A^\pm_L)_\varphi$ and $(A^\pm_R)_\varphi$. The gauge transformations are:
\begin{align}
    g_{\pm,L} = \exp(\i  \Omega_L^\pm \frac{\varphi - \Omega t}{1-\Omega} J_L^{\pm3}) \, , \qquad     g_{\pm,R} = \exp(\i  \Omega_R^\pm \frac{\varphi - \Omega t}{1+\Omega} J_R^{\pm3}) \, .
\end{align}
Therefore, the phase acquired under $\varphi \rightarrow \varphi + 2\pi$ due to the nontrivial gauge connection is
\begin{align}
\label{eq:gaugetwisting}
        g_{\pm,L}(\varphi + 2 \pi) = \exp( \frac{2 \pi \i \Omega_L^\pm}{1-\Omega} J_L^{\pm3})g_{\pm,L}(\varphi) \, , \quad     g_{\pm,R}(\varphi + 2 \pi) = \exp(  \frac{2 \pi \i \Omega_R^\pm}{1+\Omega} J_R^{\pm3})g_{\pm,R}(\varphi) \, .
\end{align}
In order to argue with have a smooth Killing spinor twisted by the above boundary conditions, we must describe the action of $J_L^{\pm3}$ and $J_R^{\pm3}$ on the Killing spinor. This is undertaken in the next subsection.

\subsection*{$SU(2)$ representations and algebra of Killing spinors}
The finite temperature black hole solution has been shown to possess a locally well defined Killing spinor \eqref{eq:KSfiniteT} with the correct boundary condition for the contractible cycle. We would like to show that this spinor has the correct $SU(2)_+ \times SU(2)_-$ representations, which we use to find the compatibility of the explicit phase \eqref{eq:KSnoncontphase} with the phase acquired due to the nontrivial gauge transformation \eqref{eq:gaugetwisting}. 

The action of the $J_L^{\pm3}$ and $J_R^{\pm3}$ on the Killing spinor is related to the algebra of preserved supersymmetries. The Killing spinor on it's own does not have any algebra, and in principle to construct the algebra one needs to construct an ADM like analog of the supercharge, as discussed in \cite{Gauntlett:1998kc,Gauntlett:1998fz,Figueroa-OFarrill:1999klq,Ortin:2015hya}. As discussed there, a proxy of this computation is to associate fermion bilinears (which are automatically Killing vectors) to anticommutators schematically as
\begin{align}
    (\bar{\epsilon}\Gamma^\mu \epsilon) \partial_\mu \sim K \, \longleftrightarrow \, \{ F, \bar{F} \} \sim B \, ,
\end{align}
where $\epsilon$ and $K$ are Killing spinors and vectors, while $F$ and $B$ are fermionic and bosonic generators of the symmetry algebra. For commutators involving bosonic and fermionic generators, the recipe in terms of Killing vectors and spinors is to measure the charge of a spinor with the Lie derivative along the direction of a Killing vector \cite{Ortin:2015hya}:
\begin{align}
    [K,F] \sim F' \, \longleftrightarrow \mathcal{L}_K \epsilon \sim \epsilon' \, ,
\end{align}
where the Lie derivative along $K$ acting on a spinor is
\begin{align}
\label{eq:spinorLieD}
    \mathcal{L}_K \epsilon = K^\mu D_\mu \epsilon + \frac14 D_{[a}K_{b]}\Gamma^{ab}\epsilon \, ,
\end{align}
using the usual spinor covariant derivative $D_\mu$. 

For the $SU(2)^L_+ \times SU(2)^L_- \times SU(2)^R_+ \times SU(2)^R_-$ isometries, a basis of Killing vector fields for the metric \eqref{eq:susyBTZmetric} in the coordinates \eqref{eq:susyBTZrotcords} in which the $S^3 \times S^3$ metric is round is given by
\begin{align}
    J^{1,+}_L &= \frac12(\cos(\tilde{\phi}_+ + \tilde{\psi}_+)\partial_{\theta_+} + \sin(\tilde{\phi}_+ + \tilde{\psi}_+) \tan(\theta_+) \partial_{\tilde{\psi}_+} - \sin(\tilde{\phi}_+ + \tilde{\psi}_+)\cot(\theta_+) \partial_{\tilde{\phi}_+} )\, , \\
    J^{2,+}_L &= \frac12(\sin(\tilde{\phi}_+ + \tilde{\psi}_+)\partial_{\theta_+} - \cos(\tilde{\phi}_+ + \tilde{\psi}_+) \tan(\theta_+) \partial_{\tilde{\psi}_+} + \cos(\tilde{\phi}_+ + \tilde{\psi}_+)\cot(\theta_+) \partial_{\tilde{\phi}_+} )\, , \\
    J^{3,+}_L &= \frac12(\partial_{\tilde{\psi}_+} + \partial_{\tilde{\phi}_+}) \, , \\
    J^{1,+}_R &= \frac12(-\cos(\tilde{\phi}_+ - \tilde{\psi}_+)\partial_{\theta_+} + \sin(\tilde{\phi}_+ - \tilde{\psi}_+) \tan(\theta_+) \partial_{\tilde{\psi}_+} + \sin(\tilde{\phi}_+ - \tilde{\psi}_+)\cot(\theta_+) \partial_{\tilde{\phi}_+} )\, , \\
    J^{2,+}_R &= \frac12(-\sin(\tilde{\phi}_+ - \tilde{\psi}_+)\partial_{\theta_+} - \cos(\tilde{\phi}_+ - \tilde{\psi}_+) \tan(\theta_+) \partial_{\tilde{\psi}_+} - \cos(\tilde{\phi}_+ - \tilde{\psi}_+)\cot(\theta_+) \partial_{\tilde{\phi}_+} )\, , \\
    J^{3,+}_R &= \frac12(\partial_{\tilde{\psi}_+} - \partial_{\tilde{\phi}_+}) \, ,
\end{align}
with a similar set of generators obtained by replacing $+ \leftrightarrow -$. These generators satisfy $D_{(a}K_{b)} =0$ in frame coordinates as well as the algebra (taking $I,J =1,2,3$):
\begin{align}
    [J^{I,\pm}_L,J^{J,\pm}_L] &= - \epsilon^{IJK} J^{K,\pm}_L \, , \\
    [J^{I,\pm}_R,J^{J,\pm}_R] &=  \epsilon^{IJK} J^{K,\pm}_R \, .
\end{align}

To find the action of these generators on the Killing spinor, we explicitly compute \eqref{eq:spinorLieD} with the understanding that the spinor \eqref{eq:KSfiniteT} is independent of all angles on the spheres in these coordinates. One obtains
\begin{align}
    \mathcal{L}_{(J^{1,+}_L)} \varepsilon^i &= - \frac{\i}{2} \Gamma_{45}\varepsilon^i \, , \qquad \mathcal{L}_{(J^{1,+}_R)} \varepsilon^i = 0 \, , \\
    \mathcal{L}_{(J^{2,+}_L)} \varepsilon^i &= - \frac{\i}{2} \Gamma_{35}\varepsilon^i \, , \qquad \mathcal{L}_{(J^{2,+}_R)} \varepsilon^i = 0  \, , \\
    \mathcal{L}_{(J^{3,+}_L)} \varepsilon^i &= -\frac{\i}{2} \Gamma_{34}\varepsilon^i \, , \qquad \mathcal{L}_{(J^{3,+}_R)} \varepsilon^i = 0 \, .
\end{align}
A similar structure is found for the $-$ derivatives,
\begin{align}
    \mathcal{L}_{(J^{1,-}_L)} \varepsilon^i &= - \frac{\i}{2} \Gamma_{78}\varepsilon^i \, , \qquad \mathcal{L}_{(J^{1,-}_R)} \varepsilon^i = 0 \, , \\
    \mathcal{L}_{(J^{2,-}_L)} \varepsilon^i &= - \frac{\i}{2}\Gamma_{68}\varepsilon^i \, , \qquad \mathcal{L}_{(J^{2,-}_R)} \varepsilon^i = 0  \, , \\
    \mathcal{L}_{(J^{3,-}_L)} \varepsilon^i &= - \frac{\i}{2} \Gamma_{67}\varepsilon^i \, , \qquad \mathcal{L}_{(J^{3,-}_R)} \varepsilon^i = 0 \, .
\end{align}
In order to reproduce the exact phase in \eqref{eq:KSnoncontphase} given the twisting \eqref{eq:gaugetwisting}, we needed to set some components of the more general Killing spinor \eqref{eq:KSfiniteT} to zero. It is easier to see from the above actions of $J^{3,\pm}_R$ that the phase associated to $g_{\pm,R}(\varphi + 2 \pi)$ is trivial for any value of $\Omega^\pm_R$ when acting on \eqref{eq:KSfiniteT}. This is is line with our expectation below \eqref{eq:killingspinorboundaryconds} that there is no constraint from supersymmetry on the right moving potentials. So the total phase around the non-contractible cycle is
\begin{align}
    g_{L}(\varphi + 2 \pi) = \exp \left( \frac{2 \pi \i \Omega_L^+}{1-\Omega} J_L^{+3} +  \frac{2 \pi \i \Omega_L^-}{1-\Omega} J_L^{-3} \right )g_{L}(\varphi)
\end{align}

Using the supersymmetry condition \eqref{eq:killingspinorboundaryconds}, $\Omega_L^- = \Omega_L^+ + \frac{2 \pi \i}{\beta}$, which implements the complex angular velocity of the metric, the above phase splits into a $\beta$ dependent part and a $\beta$ independent part. Recall we are trying to see if failure of the local Killing spinor to be periodic, \eqref{eq:KSnoncontphase}, can be canceled by the correction due to the nontrivial gauge field above. For the $\beta$ indepdent, but $\Omega_L^{+}$ dependent part, we have simply
\begin{align}
     \exp\left( \frac{2 \pi \i \Omega_L^+}{1-\Omega} (J_L^{3+} + J_L^{3-}) \right)\varepsilon^i  = \varepsilon^i \, .
\end{align}
For generic $\Omega_L$, there are no nontrivial solutions, so the spinor must be annhilated by this combination of angular momenta. Using the explicit form of the $\SU(2)$ generators in terms of Gamma matrices, we can write the above as a projector equation:
\begin{align}
\label{eq:gamma3467proj}
    (\Gamma_{34} + \Gamma_{67})\varepsilon^i = 0 \, .
\end{align}
Similar to our discussion surrounding \eqref{eq:10dprojector}, this condition is equivalent to the eigenspinor equation:
\begin{align}
    \Gamma_\Omega \varepsilon^i = \varepsilon^i \, , \qquad \Gamma_\Omega \equiv \Gamma_{3467} \, .
\end{align}
This projector commutes with those in \eqref{eq:SigmaGammaEigen} so we can impose the simultaneous conditions. This justifies the introduction of the constant spinors \eqref{eq:genconstantspinor} which were then constrained to the form \eqref{eq:constrainedconstantspinor}. 

We finally must address the $\beta$ dependent part of the phase, which reads
\begin{align}
\label{eq:betadependentphase}
    \exp \left(  \frac{-4 \pi^2}{\beta(1-\Omega)} J_L^{-3} \right )\varepsilon^i =    \exp \left(  \frac{2 \pi^2 \i }{\beta(1-\Omega)} \Gamma_{67} \right )\varepsilon^i \, .
\end{align}
Acting with this generator on the Killing spinor \eqref{eq:KSfiniteT}, where we have diagonalized the action of $\Gamma_{67}$:
\begin{align}
    \Gamma_{67} \chi_{++-} = -\i \chi_{++-} \, , \quad \Gamma_{67} \chi_{+-+} = \i \chi_{+-+} \, , \quad \Gamma_{67} \chi_{-+-} = - \i \chi_{-+-} \, , \quad \Gamma_{67} \chi_{--+} = \i \chi_{--+} 
\end{align}
We can see that \eqref{eq:betadependentphase} acting on \eqref{eq:KSfiniteT} reproduces the failure of periodicity in \eqref{eq:KSnoncontphase}. The form of the solution \eqref{eq:KSfiniteT} with two free parameters appears to be the most general one which satisfies the projector conditions and is smooth for the positive/negative frequency components. We conclude we have found a smooth supersymmetric solution by combining \eqref{eq:KSfiniteT} with the conditions (suppressing indices):
\begin{align}
\label{eq:SigmaGammaOmegaEigen}
    \sigma_1 \varepsilon = \varepsilon \, , \qquad \Gamma \varepsilon =  \varepsilon \, , \qquad \Gamma_\Omega \varepsilon = \varepsilon \, .
\end{align}
Of the 32 supercharges we started with, each of the projectors above reduces the number of free components by $\frac12$. Naively, this results in 4 real supercharges which could be paired up to the $Q$'s of \eqref{eq:extralargeRR}, ignoring the $\tilde{Q}$'s which linearize the algebra. However, this is too fast, as we can see that our spinor is not a bifundamental under $SU(2)_L^+ \times SU(2)_L^-$; the smooth solutions transform under only the diagonal, \eqref{eq:gamma3467proj}. Furthermore, we found that smoothness for the complex velocity in \ref{eq:betadependentphase} removed an additional two degrees of freedom. One can trace this back to the fact that our complex solution is computing the elliptic genus, so we must really pick a preserved $\mathcal{N}=2$ sub-algebra from the full symmetry algebra. This linearized $\mathcal{N}=2$ subalgebra means we should really find 2 $Q$'s and 2 $\tilde{Q}$'s. To fully see if this is true for the Killing spinor we found, one should really compute all the ADM charges as well as their algebra, but we will not attempt that here. 

This previous point deserves a few more comments. If we were only considering the vacuum $AdS_3 \times S^3 \times S^3 \times S^1$ solution, we would find a linearized algebra of isometries obtained from the metric and supersymmetry transformations in the standard way. But in passing to the Ramond sector to describe the black holes, the large $\mathcal{N}=4$ symmetry seems to require a nonlinear algebra and nonlinear BPS bound. The additional $\tilde{Q}$ generators and their superpartners linearize the algebra, but it was not totally clear to us how to obtain these directly from 10-dimensional supergravity. 

In any event, we have demonstrated the complex solution has the correct non-holomorphic action appropriate for the elliptic genus which was predicted by the Schwarzian theory. This means that even though the linear vs non-linear algebra of charges is somewhat obscured, the nonlinear BPS bound is correctly captured by this solution, and it permits smooth well-defined Killing spinors with the correct boundary conditions. This seems to be a rather special solution in which a large amount of supersymmetry is preserved (seemingly a full $\mathcal{N}=2$ subalgebra), all while remaining at finite temperature due to the imaginary part of the angular velocity. It would be interesting to understand if there are other finite temperature supersymmetric solutions realizing other large or nonlinear supersymmetry algebras. In the concluding section, we comment on some examples which may be studied in the Schwarzian limit, but we are unaware of any corresponding complex black hole.

\section{Conclusions and future directions}\label{sec:conc}

We studied the implications of quantum gravity corrections to spectrum of black holes in $AdS_3 \times S^3 \times S^3 \times S^1$ with large $\mathcal{N}=4$ supersymmetry. The BPS spectrum we find is a robust low energy prediction, and is much more exotic than previous results for small $\mathcal{N}=4$  near BPS black holes \cite{Heydeman:2020hhw}, as well as the large $\mathcal{N}=4$ NS sector states \cite{Eberhardt:2017fsi}. It would be very interesting to understand this BPS spectrum from the point of view of the dual SCFT, since we are making a (large $N$) prediction for a protected quantity. It is possible that the recent description in \cite{Witten:2024yod} may be useful in this regard. 

We also found non-extremal BPS black hole solutions that are asymptotically $AdS_3 \times S^3 \times S^3 \times S^1$ and have a temperature-dependent index. It would be interesting to construct asymptotically flat 10d solutions with this property, since this does not seem to happen in any known example. This temperature dependence can be traced back to the non-linear BPS relations, so a complex solution with these properties would also be helpful in understanding when and how non-linear BPS relations can arise. 

Finally, we mentioned that the large $\mathcal{N}=4$ Schwarzian is also relevant to 6d black holes that have no intermediate Virasoro symmetry. We conclude with a short overview of these backgrounds and some open questions.

\subsection{Black holes with $AdS_2 \times S^2 \times S^2 \times T^5$ throat}\label{sec:6dBH}
In addition to the $AdS_3 \times S^3 \times S^3 \times S^1$ background analyzed in the previous sections, there exists another supergravity solution which possesses a large $\mathcal{N}=4$ superconformal symmetry, this time realized as the near horizon backreacted geometry of intersecting M2 and M5 branes which gives $AdS_2 \times S^2 \times S^2 \times T^5$~\cite{Boonstra:1998yu}. Our interest in this system comes from the fact that the near-horizon region contains an AdS$_2$ factor with the large $\mathcal{N}=4$ symmetry, explicitly \emph{without} an embedding in AdS$_3$. Therefore, the usual arguments about CFT$_2$, including modular invariance, cannot be used to predict the asymptotic degeneracies of high energy states. To the best of our knowledge, there is also not a microscopic analysis, which means the Schwarzian theory seems to be the only tool for analyzing the spectrum of large $N$ microstates beyond the semiclassical Hawking formula.

In this subsection, we offer only a few comments on this solution and speculate on the applicability of our Schwarzian results to this system. We leave a full analysis of this background for future work. Most of our explanation is taken directly from \cite{Boonstra:1998yu}, but we will emphasize a few features necessary to apply our Schwarzian solution.

The microscopic brane construction in M-theory is given by the following table:
\begin{align}
\begin{array}{|c|c|c|c|c|c|c|c|c|c|c|c|}
\hline
   & 0 & 1 & 2 & 3 & 4 & 5 & 6 & 7 & 8 & 9 & 10 \\
\hline
M2_1 & \text{—} & \text{—} & \text{—} &  &  &  &  &  &  &  &  \\
\hline
M2_2 & \text{—} &  &  & \text{—} & \text{—} &  &  &  &  &  &  \\
\hline
M5_1 & \text{—} &  & \text{—} &  & \text{—} & \text{—} & \text{—} & \text{—} &  &  &  \\
\hline
M5_2 & \text{—} & \text{—} &  & \text{—} &  & \text{—} & \text{—} & \text{—} &  &  &  \\
\hline
M5_3 & \text{—} &  & \text{—} & \text{—} &  &  &  &  & \text{—} & \text{—} & \text{—} \\
\hline
M5_4 & \text{—} & \text{—} &  &  & \text{—} &  &  &  & \text{—} & \text{—} & \text{—} \\
\hline
\end{array}\label{eq:tableMBRANE}
\end{align}

The supergravity solution is obtained from taking the standard superposition of harmonic functions, then going to the near horizon limit, resulting in $AdS_2 \times S^2 \times S^2 \times R^5$. While the brane configuration is defined in terms of a noncompact asymptotically flat spacetime, once we take the near horizon limit it is possible to compactify the $x^1 \dots x^{5}$ directions to $T^5$, so we may label them as $\theta_1 \dots \theta_5$ with $\theta_i \sim \theta_i + 2\pi$. Here, we will take the $T^5$ to be a square torus with side length $2 \pi L$. This means the AdS$_2 \times S^2 \times S^2$ part of the geometry comprising $(x^0, x^6\dots x^{10})$  is also a solution of six-dimensional maximal supergravity. Unfortunately, we are not aware of any nonextremal or finite temperature solution that is asymptotically flat in six dimensions; however, see the discussion in \cite{Boonstra:1998yu} for more details. The Schwarzian analysis really only makes sense if we may turn on a small finite temperature; therefore, we will regard our discussion in this section as somewhat preliminary.

The near horizon supergravity solution is
\begin{align}
    ds^2 &= -r^2 \left(\frac{1}{R_+^2} + \frac{1}{R_-^2} \right)\d t^2 + \left(\frac{1}{R_+^2} + \frac{1}{R_-^2} \right)^{-1}\frac{\d r^2}{r^2} \nonumber\\
    &+ R_+^2(\d\theta_+^2 + \sin^2(\theta_+)\d\phi_+^2)+ R_-^2(\d\theta_-^2 + \sin^2(\theta_-)\d\phi_-^2) + L^2 \d\theta_i \d\theta^i \, , \\
    F_4 &= L^2 ( \frac{\sqrt{R_+^2 + R_-^2}}{R_+ R_-} \d t \wedge \d r \wedge \d\theta_1 \wedge \d\theta_2 +  \frac{\sqrt{R_+^2 + R_-^2}}{R_+ R_-} \d t \wedge \d r \wedge \d\theta_3 \wedge \d\theta_4 \nonumber  \\
    &  + R_+ \sin(\theta_+) \d\theta_+ \wedge \d \phi_+ \wedge \d\theta_1 \wedge \d\theta_3 - R_+ \sin(\theta_+) \d\theta_+ \wedge \d \phi_+ \wedge \d\theta_2 \wedge \d\theta_4 \nonumber  \\
    & + R_- \sin(\theta_-) \d\theta_- \wedge \d \phi_- \wedge \d\theta_1 \wedge \d\theta_4 + R_- \sin(\theta_-) \d\theta_- \wedge \d \phi_- \wedge \d\theta_2 \wedge \d\theta_3 )\, ,
\end{align}
with $i=6\dots 10$. Importantly, this background is really supported by 6 different flux terms matching the branes in \eqref{eq:tableMBRANE}. Note also that there is a minus sign in one of the $R_+$ terms to ensure supersymmetry, and also the $\theta_5$ coordinate does not enter the fluxes. 

Reference \cite{Boonstra:1998yu} argues that the isometries of this solution contain the superconformal group $\text{D}(2,1|\upalpha)$ and therefore the effective dynamics should be captured by the large $\mathcal{N}=4$ Schwarzian theory. To uniquely nail down the effective action we need to know the Schwarzian coupling $\Phi_r$. This parameter is related to the way the $AdS_2$ throat is glued to the asymptotically flat six (or potentially higher dimensional) spacetime. Since we don't have this solution away from extremality, we cannot convincingly determine this parameter, and therefore we cannot determine the gap between the BPS states and the first excited non-BPS continuum. 

Even without knowing the gap, we can still derive interesting properties of the BPS spectrum solely from the extremal solution. In particular, we can derive how the BPS black holes distribute themselves according to the angular momenta $(j_+,j_-)$ which corresponds to rotation along the first and second $S^2$ factors, respectively. This is given by $N_{j_+,j_-}$ given in \eqref{eqn:NBPS} with $R_{\text{BPS}}$ given in \eqref{eq:BPSrangeD21}. To apply these results we only need to know $S_0$ and $\upalpha$ and they are both determined by the extremal solution. The former is determined by the area of the horizon in 11d Planck units
\beq
S_0 = \frac{(2\pi)^7 R_+^2 R_-^2 L^5}{G_N^{11}} = \frac{8 R_+^2 R_-^2 L^5}{\ell_P^9} \, ,
\eeq
where we introduced the conventional eleven dimensional Newton constant and Planck length as $32 \pi^2 G_N^{11} = (2\pi \ell_P)^9$. The parameter $\upalpha$ can be inferred from the algebra between the Killing spinors, since it appears explicitly in the (anti-)commutation relations of $\text{D}(2,1|\upalpha)$. Based on Appendix A of \cite{Boonstra:1998yu}, and in particular equation 42, we expect
\beq
\upalpha = \frac{R_-^2}{R_+^2},
\eeq
which is completely analogous to its $AdS_3 \times S^3 \times S^3 \times S^1$ counterpart in type II string theory. Therefore we automatically have an interesting prediction for the spectrum of BPS states of this theory and the particular quantum mechanics living on the intersection of the M-theory branes in \eqref{eq:tableMBRANE}.

\paragraph{Acknowledgements} It is a pleasure to thank Lorenz Eberhardt, Matthias Gaberdiel, Mukund Rangamani, and Edward Witten for valuable discussions. We would like to thank especially Mukund Rangamani for numerous illuminating discussions on the 10d background and its large $\mathcal{N}=4$ Virasoro symmetry. MTH is supported by Harvard University and the Black Hole Initiative, funded in part by the Gordon and Betty Moore Foundation (Grant 8273.01) and the John Templeton Foundation (Grant 62286). XS and GJT are supported by the University of Washington and the DOE award DE-SC0011637.

\appendix

\section{$\SU(2)$ modes from $AdS_3 \times S^3 \times S^3 \times S^1$}\label{app:SU2DER}

We give an efficient derivation of the $\SU(2)$ Chern-Simons terms that arise from gravity on the $AdS_3 \times S^3 \times S^3 \times S^1$ background. We follow the approach developed by Hansen and Kraus in \cite{Hansen:2006wu}. We focus on the gauge fields that arise from the isometries of $S^3 \times S^3$ that give rise to $\SO(4) \times \SO(4)$, or equivalently, to four $\SU(2)$ gauge symmetries.

To compute the action of the gauge field $A^{ij}_\pm$ we need to determine whether it affects other fields of the background. As explained by Hansen and Kraus, it is not nontrivial to guarantee that $F_3$ will be both closed, and obtained from a local $C_2$. Applying their methods to our problem implies we need to update the RR field to
\beq
F_3 = Q_5^+ (e_3^+-\chi_3^+) + Q_5^- (e_3^--\chi_3^-) + Q_1 *(\epsilon_+ \wedge \epsilon_- \wedge \d \theta),
\eeq
where, for each three-sphere, we define (we omit the $\pm$ index to avoid cluttering)
\beq
e_3 = \frac{1}{(2\pi)^2} \epsilon_{ijkl} \left( \frac{1}{3} Dy^i Dy^j Dy^k - \frac{1}{2} F^{ij} Dy^k \right) y^l,~~~2\pi \chi_3 =  {\rm CS}(A_R)- {\rm CS}(A_L).
\eeq
where we define $D y^i = \d y^i - A^{ij} y^j$, $F^{ij}=[D,D]^{ij}$, and ${\rm CS}(A) =\frac{1}{4\pi} \operatorname{Tr}(A \d A + \frac{2}{3} A^3)$ is the integrand of the Chern-Simons action. ($\chi_3$ is the corresponding Euler class of each sphere bundle.) The first term is normalized such that $\int_{S^3} e_3 = 1$, and it is a generalization of the sphere volume form that is gauge invariant. The second term in the ansatz guarantees that the 3-form flux across the spheres is not modified, that $F_3$ is closed, and that $F_3 = \d C_2$ has a local solution for $C_2$ (whose explicit form is known but we will not need). 

The $\SU(2)$ CS terms in the action arise precisely due to the improvement terms needed in the RR 3-form. We can identify the coefficient of the CS action by analyzing the variation of the 10d action under a gauge transformation. Since $e_3$ is manifestly gauge invariant, the gauge variation of the RR form is $ \delta_\Lambda F_3 = - Q_5^+ \d \chi_2^+ - Q_5^- \d \chi_2^-$. Here we used the fact that although $\chi_3$ is not gauge invariant, the variation is a total derivative and $\delta_\Lambda \chi_3 = - \d \chi_2$ defines the 2-form $\chi_2$ (the explicit form will not be important). The variation of the ten-dimensional action is
\bea
\delta_\Lambda I &=& 2 \pi Q_5^+ \int_{\partial M_{10}} * F_3 \wedge \chi_2^+  +2 \pi Q_5^- \int_{\partial M_{10}} * F_3 \wedge \chi_2^- \nonumber\\
&=& 2 \pi Q_5^+Q_1 \int_{\partial AdS_3}  \chi_2^+  +2 \pi Q_5^- Q_1 \int_{\partial AdS_3} \chi_2^-
\ea
In the second line we integrated over the $S^3 \times S^3 \times S^1$ directions. Upon dimensional reduction to $AdS_3$ this variation is reproduced by
\beq
I \supset  -\i k^+  \int {\rm CS}(A_L^+)+\i k^+  \int {\rm CS}(A_R^+) -\i k^-  \int {\rm CS}(A_L^-)+\i k^-  \int {\rm CS}(A_R^-).
\eeq
One can show that these terms, supplemented with a careful treatment of boundary conditions, lead to two left-moving (coming in our conventions from $A_L^\pm$) and right-moving (coming in our conventions from $A_R^\pm$) $\SU(2)$ Kac-Moody algebras with levels
\beq
k^+ = Q_1 Q_5^+ ,~~~~~k^- = Q_1 Q_5^-.
\eeq
The analysis of Hansen and Kraus also shows that solutions that are pure large gauge transformations can also carry angular momentum due to these CS terms. 

\section{Path integral of large $\mathcal{N}=4$ supergravity on $AdS_3$}\label{app:N4Virasoro}

In this Appendix we derive the gravitational path integral over the charged BTZ black hole in large $\mathcal{N}=4$ supergravity on $AdS_3$.\footnote{We would like to thank Mukund Rangamani for multiple discussions regarding the content of this section.} We will begin with the theory associated to $\tilde{A}_\gamma$, which can be obtained from a Chern-Simons construction \cite{Henneaux:1999ib} and in the end upgrade it to the $A_\gamma$ theory by including the extra fields. To match the near-extremal behavior, we follow the approach taken for $\mathcal{N}=0$ in \cite{Ghosh:2019rcj} although the case considered here is considerably more complicated.

Following Maloney and Witten \cite{Maloney:2007ud} we know that the path integral over the BTZ black hole is given by the large $\mathcal{N}=4$ Virasoro vacuum character on the torus. In the case with large $\mathcal{N}=4$ supergravity this includes Virasoro descendants as well as the $\SU(2)$ ones and the fermionic ones. It also involves a sum over saddles which can be interpreted as integral spectral flow form the boundary point of view.  

Let us begin by considering the path integral on $AdS_3$ in the NS sector, transforming it to the BTZ answer via an S-move later. We also consider an insertion of $(-1)^{\sf F}$ in the partition function, implying fermions are periodic around the thermal circle in the AdS presentation (which will imply periodic fermions aroung the spatial circle when transformed to BTZ). The gravitational path integral is given by the vacuum character which was derived by Petersen and Taormina in \cite{Petersen:1989zz,Petersen:1989pp}. We refer the reader to its derivation in the original papers, their result for the character is 
\bea
Z_{AdS}&=&  q^{-\frac{\tilde{c}}{24}} \frac{\prod_{n=1}^\infty (1-z_+ z_- q^{n-\frac{1}{2}})(1-z_+^{-1} z_-^{-1} q^{n-\frac{1}{2}})(1-z_+^{-1} z_- q^{n-\frac{1}{2}})(1-z_+ z_-^{-1} q^{n-\frac{1}{2}})}{\prod_{n=1}^\infty (1-q^n)(1-z_+^2 q^n)(1-z_+^{-2} q^n) \prod_{n=1}^\infty (1-q^n)(1-z_-^2 q^n)(1-z_-^{-2} q^n) }\nonumber\\
&&\frac{1}{(z_+-z_+^{-1}) (z_--z_-^{-1})}\frac{1}{ \prod_{n=1}^\infty (1-q^n)}\nonumber\\
&& \sum_{n_\pm\in \mathbb{Z}} \sum_{\epsilon_\pm = \pm1} q^{n_+^2 k^+ + n_-^2 k^-} q^{n_++n_-}z_+^{2 \epsilon_+ n_+ k^+}z_-^{2 \epsilon_- n_- k^-}\frac{\epsilon_+ \epsilon_- z_+^{\epsilon_+} z_-^{\epsilon_-} }{1-q^{n_++n_-+\frac{1}{2}} z_+^{\epsilon_+} z_-^{\epsilon_-} }
\ea
This is the expression of Petersen and Taormina obtained by summing over large $\mathcal{N}=4$ descendants. Its interpretation as a gravitational path integral is not manifest in this presentation. We begin by making this connection more evident by rewriting this in a more illuminating form.

First, we shift $n_+ \to \epsilon_+ n_+$ and $n_- \to \epsilon_- n_-$, and then carry out the sum over $\epsilon_+$ and $\epsilon_-$. It is important to do this in this order. The partition function becomes
\bea
Z_{AdS}&=&   \frac{\prod_{n=1}^\infty (1-z_+ z_- q^{n-\frac{1}{2}})(1-z_+^{-1} z_-^{-1} q^{n-\frac{1}{2}})(1-z_+^{-1} z_- q^{n-\frac{1}{2}})(1-z_+ z_-^{-1} q^{n-\frac{1}{2}})}{\prod_{n=1}^\infty (1-q^n)(1-z_+^2 q^n)(1-z_+^{-2} q^n) \prod_{n=1}^\infty (1-q^n)(1-z_-^2 q^n)(1-z_-^{-2} q^n) }\nonumber\\
&&\frac{1}{(z_+-z_+^{-1}) (z_--z_-^{-1})}\frac{1}{ \prod_{n=1}^\infty (1-q^n)}\nonumber\\
&& \sum_{n_\pm \in \mathbb{Z}} q^{-\frac{\tilde{c}}{24}+n_+^2 k^+ + n_-^2 k^-} z_+^{2 n_+ k^+}z_-^{2 n_- k^-}\frac{(1-q)(1-z_+^2 q^{2n_+})(1-z_-^{2} q^{2n_-})}{z_+ q^{n_+} z_- q^{n_-}\prod_{\epsilon_\pm} (1-z_+^{\epsilon_+} z_-^{\epsilon_-} q^{\epsilon_+ n_+ + \epsilon_- n_- + \frac{1}{2}})} \nonumber
\ea
This is not yet in a recognizable form. Consider the following identities
\beq
\prod_{n=1}(1+z q^{m+n-1/2})(1+z^{-1} q^{-m+n-1/2}) = \frac{1}{q^{m^2/2}z^{m}}\prod_{n=1}(1+z q^{n-1/2})(1+z^{-1} q^{n-1/2}),\label{eq:idffff}
\eeq
and
\beq
\prod_{n=1}^\infty (1-z^2 q^{2m+n})(1-z^{-2} q^{-2m+n}) = \frac{(z-z^{-1}) \prod_{n=1}^\infty (1-z^2 q^n)(1-z^{-2} q^n)  }{q^{2m^2} z^{4m}(z q^m-z^{-1}q^{-m})},
\eeq
We can apply these relations to rewrite the whole prefactor as a simple function of $q$ and $z_\pm$ under the shifts $z_\pm \to z_{\pm} q^{n_\pm}$. Namely, the partition function becomes
\beq
Z_{AdS} = \sum_{n_-,n_+} q^{-\frac{\tilde{c}}{24} + \tilde{k}^+ n_+^2 + \tilde{k}^- n_-^2 } z_+^{2 \tilde{k}^+ n_+} z_-^{2 \tilde{k}^- n_-} Z_{\rm quant.}^{AdS}(q, z_+ q^{n_+}, z_- q^{n_-}), 
\eeq
where $\tilde{k}^\pm = k^\pm -1$ is the level of the $\SU(2)$ Kac-Moody algebras $\tilde{A}_\gamma$ in terms of the levels of the larger algebra $A_\gamma$, and the prefactor is
\bea
Z_{\rm quant.}^{AdS} &=& \underbrace{\frac{1}{\prod_{n=2}^\infty (1-q^n)}}_{\rm graviton}\times  \underbrace{\frac{1}{\prod_{n=1}^\infty(1-q^n) (1-z_+^2 q^n) (1-z_+^{-2} q^n)}}_{\SU(2)_+}\nonumber\\
&&\hspace{-1cm}\times \underbrace{\frac{1}{\prod_{n=1}^\infty(1-q^n) (1-z_-^2 q^n) (1-z_-^{-2} q^n)}}_{\SU(2)_-}\nonumber\\
&&\hspace{-1cm}\times \underbrace{\prod_{n=2}^\infty (1-z_+ z_- q^{n-\frac{1}{2}})(1-z_+^{-1} z_-^{-1} q^{n-\frac{1}{2}})(1-z_+^{-1} z_- q^{n-\frac{1}{2}})(1-z_+ z_-^{-1} q^{n-\frac{1}{2}})}_{\text{four gravitini}}.\label{eq:1loop3d}
\ea
This expression now has a transparent gravitational interpretation. Each term can be immediately identified with (the left-moving component of) the one-loop determinant of: the graviton resumming the contributions of the Virasoro generators, notice the product starts at $n=1$ to remove the $AdS_3$ isometries, the two $\SU(2)$ Chern-Simons modes summing over Kac-Moody generators, and the four gravitini in the bifundamental of the $\SU(2)$ fields. The connection with an explicit one-loop evaluation can be done using the results of \cite{Giombi:2008vd} 
 for bosonic fields, and \cite{David:2009xg} for fermionic ones. The powers of $z_\pm$ clearly illustrate the charge of each mode with respect to the $\SU(2)$ symmetries.

Having identified the one-loop determinants, we should reproduce the on-shell action. Before doing this, it is useful to map this result to the BTZ geometry. We did not specify yet how $q$ and $z_\pm$ are related to the moduli $\tau$ and $\alpha_\pm$ associated to the boundary torus. An S-transformation can map AdS to the BTZ black hole and this amounts to a specific choice 
$$
q = e^{2\pi \i (-1/\tau)},~~~~z_+= e^{2\pi \i \alpha_+/\tau},~~~~z_-= e^{2\pi \i \alpha_-/\tau}.
$$
The partition function in terms of $\tau$ and $\alpha$ corresponding to the BTZ black hole can be written in the suggestive form
\beq
Z_{BTZ} = \sum_{n_+,n_-\in\mathbb{Z}} e^{I_{\rm class}(\tau, \alpha_++n_+,\alpha_-+n_-)} \, \cdot \, Z_{\rm quant.}^{\rm BTZ}(\tau, \alpha_++n_+,\alpha_-+n_-), 
\eeq
where the exponent is given by
\beq
I_{\rm class}(\tau, \alpha_+,\alpha_-) = \frac{\i \pi \tilde{c}}{12 \tau} - \frac{2\pi \i \tilde{k}^+ }{\tau} \alpha_+^2 - \frac{2\pi \i \tilde{k}^-}{\tau} \alpha_-^2
\eeq
This is precisely the on-shell action whose derivation can be found in \cite{Kraus:2006wn}. The one-loop determinant is simply related to the AdS one
\beq
Z_{\rm quant.}^{BTZ}(\tau, \alpha_+,\alpha_-) = Z_{\rm quant.}^{AdS}(e^{-2\pi \i/\tau}, e^{2\pi \i \alpha_+/\tau},e^{2\pi \i \alpha_-/\tau})
\eeq
The quantity $I_{\rm class}$ is the classical action of 3d gravity coupled to the $\SU(2)$ gauge fields while $Z_{\rm quan.}^{\rm BTZ}$ is the one-loop determinants. Again, the sum over integer shifts $\alpha_\pm \to \alpha_\pm + \mathbb{Z}$ is the sum over saddles distinguished by the $\SU(2)$ connections. Notice that the relevant levels here are those associated with $\tilde{A}_\gamma$ and not $A_\gamma$. 

Following \cite{Ghosh:2019rcj} we take the near-extremal limit by taking the semiclassical limit $\tilde{k}^\pm \to \infty$ with a fixed ratio and $\tau\to \infty$ scaling as $\tilde{k}$. Define the inverse temperature as $\tau = \i \beta/ \pi$ and take $\beta \sim \tilde{k} \to \infty$. One can immediately see that the infinite products in \eqref{eq:1loop3d} become precisely the products of rotation angles of the Schwarzian theory since, for example, $1-q^n =1-e^{-2\pi \i n/ \tau} \sim \frac{2\pi^2 n}{\beta} $. Alternatively, we can rewrite the products as elliptic functions and utilize their modular properties to map the $q\to 1$ behavior to its simpler $q\to0$ behavior. This leads to the useful large $\beta$ limits
\bea
\frac{1}{\prod_{n=2}^\infty (1-q^n)}  &\sim& \frac{2\pi^{5/2}}{\beta^{3/2}}  e^{-\beta/12},\\
\frac{1}{\prod_{n=1}^\infty(1-q^n) (1-z^2 q^n) (1-z^{-2} q^n)} &\sim& \frac{2\pi \alpha}{\sin 2\pi \alpha}\, \frac{\pi^{3/2}}{\beta^{3/2}} \, e^{\beta/4},\\
\prod_{n=1}^\infty (1-z q^{n-1/2})(1-z^{-1} q^{n-1/2})  &\sim& 2\, \cos \pi \alpha \, e^{-\beta/6}.
\ea
In the last line, we evaluated the gravitino one-loop determinant as if all modes contribute, including $n=1$, since we will need this later. The true one-loop determinant removing the zero-mode $n=1$ can be easily obtained from that result
\beq
\prod_{n=2}^\infty (1-z q^{n-1/2})(1-z^{-1} q^{n-1/2})  \sim \frac{\pi^4}{\beta^2}\frac{2\, \cos \pi \alpha \,}{(1-4\alpha^2)} e^{-\beta/6}
\eeq
Putting everything together we obtain the low-temperature limit of the one-loop determinant in the following form
\beq
Z^{BTZ}_{\rm quant.} \sim e^{\frac{\beta}{12} }\frac{2}{\pi^{5/2}\sqrt{\beta}} \frac{2 \alpha_+}{\sin 2\pi \alpha_+}\frac{2\alpha_-}{\sin 2\pi \alpha_-} \, \frac{2 \cos \pi (\alpha_++\alpha_-) \, 2 \cos \pi (\alpha_+-\alpha_-)}{(1-4(\alpha_++\alpha_-)^2)(1-4 (\alpha_+-\alpha_-)^2)} ,
\eeq
which automatically can be matched with \eqref{eqn:LN422}, up to a possible overall rescaling and a shift of ground-state energy. The low-temperature limit of the action can be easily obtained
\beq
I_{\rm class.} \sim \frac{\pi^2 \tilde{c}}{12\beta} - \frac{2 \pi^2 k^+}{\beta} \alpha_+^2 - \frac{2 \pi^2 k^-}{\beta} \alpha_-^2
\eeq
which allows us to identify the Schwarzian coupling and the $\SU(2)$ mode compressibilities
\beq
\Phi_r = \frac{\tilde{c}}{24}\sim \frac{k^+k^-}{4(k^++k^-)}  ,~~~~ \gamma_- = \frac{\tilde{c}}{6 k^+}\sim \frac{k^-}{k^++k^-},~~~~\gamma_+=\frac{\tilde{c}}{6k^-}\sim \frac{k^+}{k^++k^-}.
\eeq
Since we are taking large $k^\pm$ limit we can neglect the final shifts of these parameters involved in going from $A_\gamma$ to $\tilde{A}_\gamma$. These are the identifications used in Section \ref{sec:STI}. We do not currently know how to derive the exact density of states at this point (this would require deriving the modular transformation properties of these large $\mathcal{N}=4$ characters). But the fact that we show that the partition function is well approximated by the Schwarzian one at low temperatures automatically imply that we can import the Schwarzian spectrum derived in Section \ref{sec:LargeN4Review}, supporting the results in Section \ref{sec:STI} regarding their application to $AdS_3 \times S^3 \times S^3 \times S^1$ background of string theory. 

The contribution from the right-movers is an identical copy of the partition function above in terms of right-moving generators which we denote $\SU(2)^-_R \times \SU(2)^+_R$. In the near-extremal limit $\tau \sim \i \beta/ \pi $, scaling with $c$, we need to take $\bar{\tau} \to 0$. If $\bar{\tau} \to 0$ then the right moving contribution becomes well approximated by the Cardy regime. In a fixed angular momentum and $\tilde{\SU(2)}_\pm$ charges this multiplies the partition function above $Z$ by
\beq
Z_{BTZ} \to Z_{BTZ} \cdot e^{S_0},~~~S_0 = 2\pi \sqrt{\frac{c}{6}\Big(P- \frac{\bar{j}_-^2}{k^-} -\frac{ \bar{j}_+^2}{k^+}\Big)},
\eeq
and therefore simply determines the zero-temperature entropy parameter of the Schwarzian theory. Since in the right-moving sector descendants are not important, the derivation of this observation is almost identical to \cite{Ghosh:2019rcj}.

We now discuss the evaluation of the index using the gravitational path integral. Having an explicit expression for the one-loop partition function for arbitrary chemical potentials makes this analysis very transparent. We will discuss the result for the $\tilde{A}_\gamma$ theory, since we already know that the role of the $\U(1)$ multiplet is straightforward to correct by applying a version of the helicity supertrace.

The boundary conditions for the gravitational path integral are supersymmetric whenever we satisfy $\alpha_- = \alpha_+ + 1/2$. In terms of the modular parameters of the boundary, this implies $z_+= z_- \cdot \sqrt{q}$. The one-loop determinant tends to vanish because of the fermion one-loop determinant except for some special saddles. Indeed
\bea
Z_{\rm quant.}^{BTZ} &\propto& \prod_{n=2}^\infty (1-z_+^{-1}z_- q^{-n_-+n_++n-\frac{1}{2}})(1-z_+^{}z_-^{-1} q^{-n_++n_-+n-\frac{1}{2}}),\nonumber\\
&\propto& \prod_{n=2}^\infty (1-q^{n_+-n_-+n-1})(1- q^{n_--n_++n})
\ea
This product will vanish unless $n_--n_++2>0$ and $n_+-n_-+1>0$. This leaves only two possibilities
$$
n_-=n_+ ,~~~~{\rm or}~~~~n_-=n_+-1.
$$
For the reasons explained in \cite{Iliesiu:2021are}, this implies that only for those saddles the solution is supersymmetric in the bulk as well and has the potential to contribute to the index. Let us focus on one, say $n_-=n_+$, and evaluate the full one-loop determinant. After multiple cancellations between modes (as expected from supersymmetry) we obtain
\bea
Z_{\rm quant.}^{BTZ} &=&   \frac{1}{\prod_{n=1}^\infty (1-q^n) (1-z_-^{-2} q^{n})(1-z_-^2 q^{n})} \frac{z_-^{1-4n_-} q^{n_-(2n_- -1)}}{z_--z_-^{-1}} \neq 0,
\ea
and similarly for the other choice with $n_-=n_+ -1$
\bea
Z_{\rm quant.}^{BTZ} &=&  - \frac{1}{\prod_{n=1}^\infty (1-q^n) (1-z_-^{-2} q^{n})(1-z_-^2 q^{n})} \frac{z_-^{-1-4n_-} q^{n_-(2n_- +1)}}{z_--z_-^{-1}} \neq 0.
\ea
We can combine these results with the classical action and get 
\bea
Z &\propto & \sum_{n_-\in \mathbb{Z}} \sum_{\epsilon=\pm} \epsilon e^{\frac{\i \pi (\tilde{c}+6)}{12\tau} - \frac{2\pi \i k^+}{\tau} (\alpha_- +n_--\epsilon/2)^2-\frac{2\pi \i k^-}{\tau} (\alpha_- +n_-)^2}
\ea
where $k^+ =\tilde{k}^+ +1$ and $k^- = \tilde{k}^-+1$ and with an $n_-$ independent prefactor. The index as computed from 3d gravity is proportional to the index evaluated from the Schwarzian theory.

Finally, we can easily go from $\tilde{A}$ to $A$ by restoring the contribution of the $\U(1)$ multiplet. This multiplet is decoupled from the rest of the large $\mathcal{N}=4$ supergravity degrees of freedom, and therefore its effect is multiplicative. The partition function can also be found in \cite{Petersen:1989pp,Petersen:1989zz}, although it is incomplete; it lacks integral spectral flow required for a discrete charge spectrum \cite{Eguchi:2008ct}. Incorporating spectral flow the partition function is
\bea
Z_{\mathcal{S}} &=& q^{1/8} \sum_{n_0\in \mathbb{Z}}  q^{\frac{k n_0^2}{2} }z_0^{k n_0} \frac{1}{\prod_{n=1}^\infty (1-q^n)} \nonumber\\
&&\hspace{-0.1cm}\times \prod_{n=1}^\infty (1-z_+ z_- q^{n-\frac{1}{2}})(1-z_+^{-1} z_-^{-1} q^{n-\frac{1}{2}})(1-z_+^{-1} z_- q^{n-\frac{1}{2}})(1-z_+ z_-^{-1} q^{n-\frac{1}{2}})
\ea
In principle one can generalize this and sum only over $n_0 \in N\cdot \mathbb{Z}$ with $N$ some integer \cite{Eguchi:2008ct}. This would lead to fractional $\U(1)$ charges in the theory. To obtain the contribution to the black hole partition function we apply an S-move and add the term associated to the anomaly in the Kac-Moody algebra. For concreteness let us present the final answer including all fields for the supergravity associated to $A_\gamma$. The answer is  
\beq
Z_{BTZ} = \sum_{n_0\in\mathbb{Z}}\sum_{n_+,n_-\in\mathbb{Z}} e^{I_{\rm class}(\tau, \alpha_++n_+,\alpha_-+n_-,\alpha_0+n_0)} \, \cdot \, Z_{\rm quant.}^{\rm BTZ}(\tau, \alpha_++n_+,\alpha_-+n_-,\alpha_0+n_0), 
\eeq
where now
\beq
I_{\rm class}(\tau, \alpha_+,\alpha_-) = \frac{\i \pi c}{12 \tau} - \frac{2\pi \i k^+ }{\tau} \alpha_+^2 - \frac{2\pi \i k^-}{\tau} \alpha_-^2-  \frac{\pi \i (k^++k^-)}{\tau} \alpha_0^2
\eeq
Notice that in the first term we have $c = \tilde{c}+3$, the second term coming from the power $q^{1/8}$ in the extra multiplet partition function. The shift from $\tilde{k}^\pm$ to $k^\pm$ arises from the application of \eqref{eq:idffff} to the $\U(1)$ multiplet fermion partition function. Therefore we recover expressions natural from the point of view of the levels and central charge of the $A_\gamma$ algebra. The one-loop determinant of the full theory is
\bea
Z_{\rm quant.}^{BTZ} &=& \underbrace{\frac{1}{\prod_{n=2}(1-q^n)} }_{\text{graviton}} \times \underbrace{\prod_{\pm} \frac{1}{\prod_{n=1}^\infty (1-q^n) (1-z_\pm^2 q^n)(1-z_\pm^{-2} q^n)}}_{\SU(2)_-\times \SU(2)_+\text{ gauge fields}}\times \underbrace{\frac{1}{\prod_{n=1}(1-q^n)}}_{\U(1)\text{ mode} }\nonumber\\
&&\times \underbrace{\prod_{\pm\pm} \prod_{n=2}^\infty (1-z_+^{\pm1} z_-^{\pm1}  q^{n-1/2} )}_{\text{four gravitini}}\times \underbrace{\prod_{\pm\pm} \prod_{n=1}^\infty (1-z_+^{\pm1} z_-^{\pm1}  q^{n-1/2} )}_{\text{four fermions in $\U(1)$ mult.}} .\label{eq:app:1loopAdS3}
\ea
This is precisely the form of the one-loop determinant of the `linearized' large $\mathcal{N}=4$ Schwarzian theory. 

In the main text, we focus on an ensemble of fixed $\U(1)$ charges. The one-loop determinant in such an ensemble is easy to compute, and it is given by \eqref{eq:app:1loopAdS3} above after removing the contribution labeled `$\U(1)$ mode'. However, it does include the quantum fluctuations of the fermionic partners in the $\U(1)$ mode.

To recapitulate, this is the (left-moving sector of the) gravitational path integral of large $\mathcal{N}=4$ supergravity, written in a way that emphasizes its one-loop nature. This is also the vacuum character evaluated by Petersen and Taormina but incorporating $\U(1)$ integral spectral flow, necessary to avoid a continuum of charges. For more general theories, there would be further one-loop determinants from other matter fields and further corrections, although in the near-extremal limit this gives the dominant contribution.

\section{Supergravity conventions}\label{app:sugraconventions}
Here we choose an explicit representation for the 10-dimensional gamma matrices. Our choice is slightly non-standard and is somewhat adapted so it is easier to solve the various projection conditions for the Euclidean black hole. Other possible choices convenient for Killing spinors on products of spheres are found in \cite{Boonstra:1998yu}.
\begin{align}
    \Gamma_0 &= \i \sigma_1 \otimes \mathbf{1}_2 \otimes \mathbf{1}_2 \otimes \mathbf{1}_2 \otimes \mathbf{1}_2 \, \, \, \, \qquad \Gamma_5 = \sigma_3 \otimes \sigma_3 \otimes \mathbf{1}_2 \otimes \mathbf{1}_2 \otimes \mathbf{1}_2\, , \nonumber \\
    \Gamma_1 &= \sigma_2 \otimes \mathbf{1}_2 \otimes \mathbf{1}_2 \otimes \mathbf{1}_2 \otimes \mathbf{1}_2 \, \qquad \, \, \, \, \, \Gamma_6 = \sigma_3 \otimes \sigma_2 \otimes \sigma_2 \otimes \sigma_1 \otimes \mathbf{1}_2 \, , \nonumber \\
    \Gamma_2 &= \sigma_3 \otimes \sigma_1 \otimes \mathbf{1}_2 \otimes \mathbf{1}_2 \otimes \mathbf{1}_2 \, \qquad \, \, \, \, \, \Gamma_7 = \sigma_3 \otimes \sigma_2 \otimes \sigma_2 \otimes \sigma_3 \otimes \mathbf{1}_2\, , \nonumber \\
    \Gamma_3 &= -\sigma_3 \otimes \sigma_2 \otimes \sigma_3 \otimes \mathbf{1}_2 \otimes \mathbf{1}_2 \, \qquad \Gamma_8 = -\sigma_3 \otimes \sigma_2 \otimes \sigma_2 \otimes \sigma_2 \otimes \sigma_1 \, , \nonumber \\
    \Gamma_4 &= -\sigma_3 \otimes \sigma_2 \otimes \sigma_1 \otimes \mathbf{1}_2 \otimes \mathbf{1}_2 \, \qquad \Gamma_9 = -\sigma_3 \otimes \sigma_2 \otimes \sigma_2 \otimes \sigma_2 \otimes \sigma_3 \, , \nonumber \\
\end{align}
These satisfy $\{\Gamma_a, \Gamma_b \} = 2 \eta_{ab}$ in Lorentzian signature. Additionally, the Type IIB spinors are chiral with respect to the $\Gamma_{11}$ matrix, which we take to be $\Gamma_{11} = \Gamma_0 \Gamma_1 \dots \Gamma_9$. Additionally, we have
\begin{align}
    \Gamma_{34} = \mathbf{1}_2 \otimes \mathbf{1}_2 \otimes \i \sigma_2 \otimes \mathbf{1}_2 \otimes \mathbf{1}_2 \, \qquad  \Gamma_{67} = \mathbf{1}_2 \otimes \mathbf{1}_2 \otimes \mathbf{1}_2 \otimes -\i \sigma_2 \otimes \mathbf{1}_2 
\end{align}
This is useful because it diagonalizes the spin with respect to $J_L^{3;\pm}$.

We also introduced constant spinors $\chi$ which satisfy the various projection conditions:
\begin{align}
    \Gamma_{11}\chi = - \chi \, , \quad  \Gamma \chi = \chi \, , \quad \Gamma_{3467} \chi = \chi \, .
\end{align}
These can be solved in terms of the basis,
\begin{align}
    \chi = \left ( \chi_{++-} + \chi_{+-+} + \chi_{-+-} + \chi_{--+}\right ) \, ,
\end{align}
we have the explicit cumbersome expressions
\begin{align}
    \chi_{++-} &=  \frac{-1}{R_- - \i R_+ +  \sqrt{R_+^2 + R_-^2}} \left( \begin{bmatrix}
        1\\
        0
    \end{bmatrix} \otimes \begin{bmatrix}
        1 \\
        \i
    \end{bmatrix} \otimes \begin{bmatrix}
        1 \\
        \i
    \end{bmatrix} \otimes \begin{bmatrix}
        1 \\
        \i
    \end{bmatrix}\otimes \begin{bmatrix}
        1 \\
        -\i
    \end{bmatrix} \right) \, , \nonumber \\
    &+  \frac{-1}{-R_- + \i R_+ +  \sqrt{R_+^2 + R_-^2}} \left( \begin{bmatrix}
        1\\
        0
    \end{bmatrix} \otimes \begin{bmatrix}
        1 \\
        -\i
    \end{bmatrix} \otimes \begin{bmatrix}
        1 \\
        \i
    \end{bmatrix} \otimes \begin{bmatrix}
        1 \\
        \i
    \end{bmatrix}\otimes \begin{bmatrix}
        1 \\
        \i
    \end{bmatrix} \right) \, , \\
    \chi_{+-+} &=  \frac{1}{R_- - \i R_+ +  \sqrt{R_+^2 + R_-^2}} \left( \begin{bmatrix}
        1\\
        0
    \end{bmatrix} \otimes \begin{bmatrix}
        1 \\
        -\i
    \end{bmatrix} \otimes \begin{bmatrix}
        1 \\
        -\i
    \end{bmatrix} \otimes \begin{bmatrix}
        1 \\
        -\i
    \end{bmatrix}\otimes \begin{bmatrix}
        1 \\
        \i
    \end{bmatrix} \right) \, , \nonumber \\
    &+  \frac{1}{-R_- + \i R_+ +  \sqrt{R_+^2 + R_-^2}} \left( \begin{bmatrix}
        1\\
        0
    \end{bmatrix} \otimes \begin{bmatrix}
        1 \\
        \i
    \end{bmatrix} \otimes \begin{bmatrix}
        1 \\
        -\i
    \end{bmatrix} \otimes \begin{bmatrix}
        1 \\
        -\i
    \end{bmatrix}\otimes \begin{bmatrix}
        1 \\
        -\i
    \end{bmatrix} \right) \, , \\
    \chi_{-+-} &=  \frac{1}{-R_- + \i R_+ +  \sqrt{R_+^2 + R_-^2}} \left( \begin{bmatrix}
        0\\
        1
    \end{bmatrix} \otimes \begin{bmatrix}
        1 \\
        \i
    \end{bmatrix} \otimes \begin{bmatrix}
        1 \\
        \i
    \end{bmatrix} \otimes \begin{bmatrix}
        1 \\
        \i
    \end{bmatrix}\otimes \begin{bmatrix}
        1 \\
        \i
    \end{bmatrix} \right) \, , \nonumber \\
    &+  \frac{-1}{R_- - \i R_+ +  \sqrt{R_+^2 + R_-^2}} \left( \begin{bmatrix}
        0\\
        1
    \end{bmatrix} \otimes \begin{bmatrix}
        1 \\
        -\i
    \end{bmatrix} \otimes \begin{bmatrix}
        1 \\
        \i
    \end{bmatrix} \otimes \begin{bmatrix}
        1 \\
        \i
    \end{bmatrix}\otimes \begin{bmatrix}
        1 \\
        -\i
    \end{bmatrix} \right) \, , \\
    \chi_{--+} &=  \frac{1}{R_- - \i R_+ +  \sqrt{R_+^2 + R_-^2}} \left( \begin{bmatrix}
        0\\
        1
    \end{bmatrix} \otimes \begin{bmatrix}
        1 \\
        \i
    \end{bmatrix} \otimes \begin{bmatrix}
        1 \\
        -\i
    \end{bmatrix} \otimes \begin{bmatrix}
        1 \\
        -\i
    \end{bmatrix}\otimes \begin{bmatrix}
        1 \\
        \i
    \end{bmatrix} \right) \, , \nonumber \\
    &+  \frac{-1}{-R_- + \i R_+ +  \sqrt{R_+^2 + R_-^2}} \left( \begin{bmatrix}
        0\\
        1
    \end{bmatrix} \otimes \begin{bmatrix}
        1 \\
        -\i
    \end{bmatrix} \otimes \begin{bmatrix}
        1 \\
        -\i
    \end{bmatrix} \otimes \begin{bmatrix}
        1 \\
        -\i
    \end{bmatrix}\otimes \begin{bmatrix}
        1 \\
        -\i
    \end{bmatrix} \right) \, .
\end{align}

\bibliographystyle{utphys2}
{\small \bibliography{Biblio}{}}

\end{document}